\documentclass[twocolumns]{aa}
\usepackage{graphicx}
\usepackage{txfonts}
\usepackage{soul}
\usepackage{mhchem}
\usepackage{hyperref}
\makeatletter
\newcounter{chemeqn,xcolor}

\makeatother

\begin{document}

   \title{Unmasking the physical information inherent to  interstellar spectral line profiles with Machine Learning}

   \subtitle{I. Application of LTE to HCN and HNC transitions.}

\author{
Edgar Mendoza$^{\ast}$ \inst{1}  \and
Pietro Dall'Olio\inst{1,2} \and Luciene S. Coelho\inst{3} \and
Antonio Peregrín\inst{4,5} \and 
Samuel López-Domínguez\inst{6} \and
Floris F. S. van der Tak\inst{7} \and
Miguel Carvajal$^{\ast}$\inst{1,8}
}

\institute{Dept. Ciencias Integradas, Facultad de Ciencias Experimentales, Centro de Estudios Avanzados en F\'isica, Matem\'atica y Computaci\'on, Unidad Asociada GIFMAN, CSIC-UHU, Universidad de Huelva, Spain
\and 
Lyon College, Batesville (AR), U.S.
\and
Planet\'ario Juan Bernardino Marques Barrio, Instituto de Estudos Socioambientais, Universidade Federal de Goi\'as
\and
Centro de Estudios Avanzados en F\'isica, Matem\'atica y Computaci\'on, Universidad de Huelva, Spain
\and
Andalusian Research Institute in Data Science and Computational Intelligence, Universidad de Huelva, Spain
\and
Dept. Tecnologías de la Información, Escuela Técnica Superior de Ingeniería, Centro de Estudios Avanzados en F\'isica, Matem\'atica y Computaci\'on, Universidad de Huelva, Spain
\and
SRON Netherlands Institute for Space Research \& Kapteyn Astronomical Institute, University of Groningen, 9747 AD Groningen, the Netherlands
\and
Instituto Universitario Carlos I de Física teórica y computacional, Universidad de Granada, Spain
}

\titlerunning{Machine Learning estimate of the physical conditions from the interstellar spectral line profiles}
\authorrunning{Mendoza {\em et al.}}

   \date{Received --; accepted --}


  \abstract
   {Physical and chemical conditions (kinetic temperature, volume density, molecular composition, \dots) of interstellar clouds are inherent in their line spectra at mm-submm wavelengths. 
   Therefore, the spectral line profiles could be used
to estimate the physical conditions of a given source.
}
  {
We  present a new bottom-up approach, based on
Machine Learning  (ML) algorithms, to extract the physical conditions in a straightforward way from the line profiles without using radiative transfer equations.
}
   {We have simulated, for the typical physical conditions of dense molecular clouds and star-forming regions, the emission in spectral lines of the two isomers HCN and HNC, from $J=1-0$ to $J=5-4$ between 30 and 500~GHz, which are commonly  observed in dense molecular clouds and star forming regions.
The generated data cloud distribution has been parametrised using the line intensities and widths to enable a new way to analyse the spectral line profiles and to infer the physical conditions of the region. The line profile parameters have been charted to the HNC/HCN ratio and the excitation temperature of the molecule(s). 
 Three  ML algorithms have been trained, tested and compared aiming to unravel the excitation conditions of HCN and HNC and their abundance ratio.
 }
 {Machine Learning results obtained with two spectral lines, one for each isomer HCN/HNC, have been compared with the Local Thermodynamic Equilibrium (LTE) analysis for the cold source R CrA IRS 7B. The estimate of the excitation temperature and of the abundance ratio, in this case considering the two spectral lines, is in agreement with our LTE analysis. The complete optimization procedure of the algorithms (training, testing and prediction of the target quantities) have the potential to predict interstellar  cloud properties from line profile inputs at lower computational cost than before.
 }
{It is the first time that the spectral line profiles are mapped according to the physical conditions charting the ratio of two isomers and the excitation temperature of the molecules. In addition, a bottom-up approach starting from a set of simulated spectral data at different physical conditions is proposed to interpret line observations of interstellar regions and to estimate their physical conditions. This new
approach presents the potential relevance to unravel hidden interstellar conditions with the use of ML methods. 
}
 

   \maketitle
%

\section{Introduction}

The physical properties of the gas in the Interstellar Medium (ISM), including temperature, density, molecular composition, and abundance, can be determined from the characterization of the molecular spectra of infrared and radio line surveys~\citep{draine2011,shaw2006,cernicharo2023,mcguire2020}.
This characterization is achieved, first and foremost, by ongoing efforts to measure and analyse molecular spectra in the laboratory~(see, e.g., \citet{pickett1998,endres2016,carvajal2009}) and to derive the required spectroscopic quantities, i.e., line strengths, rovibrational energies, internal partition functions, \dots~\citep{carvajal2010,roueff2013,lefloch2018,mendoza2018,tak2020,mendoza2023,carvajal2019,carvajal2024}. 
Thus, from the conditions in the laboratory, assumed under control, to conditions in the 
interstellar media of galaxies, 
the spectral lines can be extrapolated allowing us to determine the physical conditions of the regions.
Nevertheless, the ISM spectral lines stack a huge amount of information about the physical conditions of the source and of the observation techniques. 
In fact, the profiles of the spectral lines observed in an ISM survey contain an information richer 
than the one intrinsic of each molecule
- coming from the quantum mechanical properties enclosed by the molecule itself - measured in the lab, i.e., the physical conditions of the surveyed objects and of the observations: excitation temperature (T$_{\mbox{ex}}$), abundance ratio, source's size ($\theta$), linewidths ($\Delta V_{\rm sys}$),  \dots~\citep{mcguire-science2021,tak2007,tak2020}. 
Therefore, in molecular astronomy and astrochemistry, the physical and molecular properties of the observed sources are inherent in each spectral line profile, although they are hidden in the quantities involved in the transition. 
Conventionally, by means of radiative transfer methods, it is possible to infer the physical properties from astronomical maps and line surveys~\citep{holdship2022}.

In this work we propose a new approach, based on Machine Learning (ML) algorithms, to extract the physical conditions in a straightforward way from the spectral line profiles, without using radiative transfer equations. 
The process followed by us is the same that was recently suggested for the application of ML in astronomy~\citep{buchner2024}.
In particular, our work addresses to determine fundamental quantities in astrochemistry as the temperatures and column densities in the ISM~\citep{draine2011,Jorgensen2020}, exploiting, with the use of ML algorithms,
the spectral properties of  molecular  lines.  Previously, \cite{Gratier2021} evaluated the use of radio molecular line observations, including CO isotopologues, HCO$^+$, N$_2$H$^+$, and CH$_3$OH, to predict H$_2$ column densities through ML methods such as Random Forest. 
\cite{Bron2021} presented a general method based on a grid of models covering the full range of possible values for unknown physical parameters, such as gas density, temperature, and depletion, to identify tracers of the ionization fraction in dense and translucent gas using Random Forest.

We here focused on the molecules hydrogen cyanide (HCN) and hydrogen isocyanide (HNC), which are among the most abundant isomers in dense molecular clouds. They play a crucial role in the chemistry of molecular clouds and are regarded as precursors of complex molecules like the nucleobase adenine \citep{Jung2013}. Although HCN is more stable than HNC because of its lower potential minimum~\citep{jamil2019,Zamir2022}, the isomerisation between HCN and HNC is a key reaction in astrochemistry, driven by quantum tunnelling and thermodynamic factors. Recent studies employing ML techniques have successfully predicted the reactivity boundaries of this isomerisation without using the classical reaction dynamics theory \citep{Yamashita2023}. 
 The interconversion between HCN and HNC can occur both in the gas phase and on icy grain surfaces under the distinct conditions of the ISM. Studies indicates that their abundances  can be influenced by various physicochemical factors (e.g., \citealt{Mendes2012,Graninger2014,Baiano2022}). Consequently, the isomeric ratios of HCN to HNC can offer valuable insights into the evolution and properties of interstellar objects.
From observations in infrared to sub-mm wavelengths, the detection of both isomers in star-forming regions at various evolutionary stages is well-documented. Recent findings on $^{12}$C/$^{13}$C ratios have been obtained from far-infrared observations of HNC and H$^{13}$CN in the hot core Orion IRc2 \citep{Nickerson2021}. Additionally, the isomeric ratios and isotopic fractionation of HCN and HNC have been discussed for their role as a chemical clock, providing insights about the chemical evolution of star-forming regions at different stages  (e.g. \citealt{Jin2015,Pazukhin2023}). Early studies of the Orion KL hot core combining observational and experimental data revealed that the HNC/HCN ratio decreases with a rising temperature and density \citep{schilke1992,Tachikawa2003}. \cite{hacar2020} analysed the intensity ratios of the $J$=1--0 lines of HCN and HNC across the Integral Shape Filament in Orion, correlating them with the gas kinetic temperature. 
Empirical linear fit calibrations were developed for regimes of low-temperature ($T_K <$ 40~K) and high-temperature ($T_K \gtrsim$ 40~K). 
HNC generally traces  cold environments (10~K), but in hot regions, its presence is believed from active ion-molecule chemistry and infrared pumping. \cite{Perez2007} discussed this scenario through the analysis of HCN and HNC 3--2 transitions in several Seyfert galaxies. In the ultra-luminous Infrared Galaxy IRAS 20551-4250, \cite{Imanishi2017} used ALMA observations of HCN, HNC and HCO$^+$ to analyse their excitation conditions and implications for infrared radiative pumping.

The study of the ISM relies heavily on the accurate modelling of the physical and chemical properties of detected molecules. In the literature, various methods have been developed to estimate physical properties using observational data and computational techniques. Among these, sampling methods—particularly Bayesian inference—have proven to be statistically robust. For instance, nested sampling has been applied to chemical and radiative transfer models \citep{behrens2022}. Sequential Monte Carlo samplers and Markov Chain Monte Carlo (MCMC) methods have been used to study the evolution of interstellar dust in the context of galaxies \citep{Galliano2021, Ramambason2022}. Additionally, MCMC has been employed to analyse spectral data from sources such as TMC-1 \citep{Gratier2016}, while sequential Monte Carlo methods have been applied in other astrophysical contexts \citep{Lebouteiller2022}. Spectroscopic data analysis has also benefited from methods such as minimum $\chi^2$ fitting \citep{Joblin2018} and gradient-based algorithms \citep{Paumard2022}. Furthermore, similar techniques have been utilized earlier~\citep{Galliano2003}. Traditionally,  Local Thermodynamic Equilibrium (LTE) models have provided a simplified, yet effective approach for characterizing excitation conditions due to their computational feasibility and broad applicability across a wide range of molecules and transitions. In this work, we present the first generalized framework for  developing and validating ML methods to predict parameters of astrochemical and astrophysical interest under LTE conditions.

The problem has been defined to derive two outcomes from the algorithms  --excitation temperatures and the HNC/HCN ratios-- using a dataset that integrates synthetic and semiempirical data, combining LTE models with observations, as detailed in subsequent sections. Given the complexities involved in data handling, we focus on the versatile case of interstellar tracers HCN and HNC to calibrate and evaluate different ML algorithms under LTE conditions. The use of LTE models allows the spectral simulations and analysis to incorporate numerous inputs and variables associated with spectroscopy and observational parameters —i.e., transition lines, rest frequencies, systemic velocities, angular and spectral resolution— while it also enables the generation of a manageable synthetic dataset for training and testing algorithms. Although this approach introduces a bias by restricting the study to LTE-based assumptions~\citep{Roueff2024} and only a subset of molecules (two out of over 300 known in the ISM), our defined problem and workflow are structured to allow future expansion of hypotheses, inclusion of additional molecular species, and adaptation to more complex data treatments.

The accuracy of predictions made by neural networks, artificial intelligence, and ML methods depends notably on the volume and quality of the data used during the training~\citep{Veneri2023,Priestley2023}. 
Thus, with the purpose of introducing this new approach, first we have simulated LTE spectral line profiles of the isomers HCN and HNC according to the ISM physical conditions across a broad range of temperatures ($T_{\rm exc}$=5--150~K), frequencies ($\nu$=30--500~GHz) and other quantities. The profiles data, fitted as Gaussian functions, have been parametrised considering the line intensities and widths, generating a data cloud distribution for an initial analysis of the data set. Then we have assessed three different ML algorithms using semiempirical lines, derived from archival data from single-dish telescopes (e.g., IRAM-30m and APEX) alongside LTE models, in the excitation conditions of the Orion KL hot core as a reference. Afterwards a realistic system such as R CrA IRS 7B, which is a cold source in the R Corona Australis star-forming region, has been tackled for testing this approach in a realistic scenario using observed spectra. For this latter case, we have estimated the excitation temperature and the HNC/HCN ratio and they have been compared with the results obtained from LTE analysis obtained in this work.
This work represents a preliminary contribution given the multiple questions that raise in the application of this new approach when the level of complexity,  related to the spectroscopy, radiative scenarios, instruments, and astronomical sources~ \citep{Floris2011,Martin2021,Nyheim2024}, increases.

This article is organized as follows: Section~\ref{sec2} describes the methodology of the proposed novel approach, detailing the ML algorithms employed, the LTE spectral models, and the data processing of the observations obtained from the ESO archive. Section~\ref{sec-results} discusses the  predictions of the approach on semiempirical data from a representative hot core. 
In addition, this approach is applied to the R CrA IRS 7B source and its results are compared with those obtained from a mainstream method. Section~\ref{sec4} draws the conclusions of this novel approach.

\section{Methodology}
\label{sec2}

In this section we explore the effect of the physical conditions on the profiles of (sub)mm lines and we make the most of this analysis to estimate the HNC/HCN ratios and the excitation temperature of a source with the use of ML algorithms. Although this study can be addressed considering any molecule, we have considered the spectral lines from $J=1-0$ to $J=5-4$ of the two isomers HCN and HNC. These two isomers were detected in many sources and have been proven suitable to determine the 
gas temperature and the evolution
of an interstellar object~\citep{hacar2020,schilke1992}. They are good candidates for a comprehensive analysis of the spectral line profiles in the study of the ISM and, in particular, when ML is going to be applied.

\begin{figure*}
  \begin{center}
    \setlength{\unitlength}{1pt} 
    \includegraphics[width = 15cm]{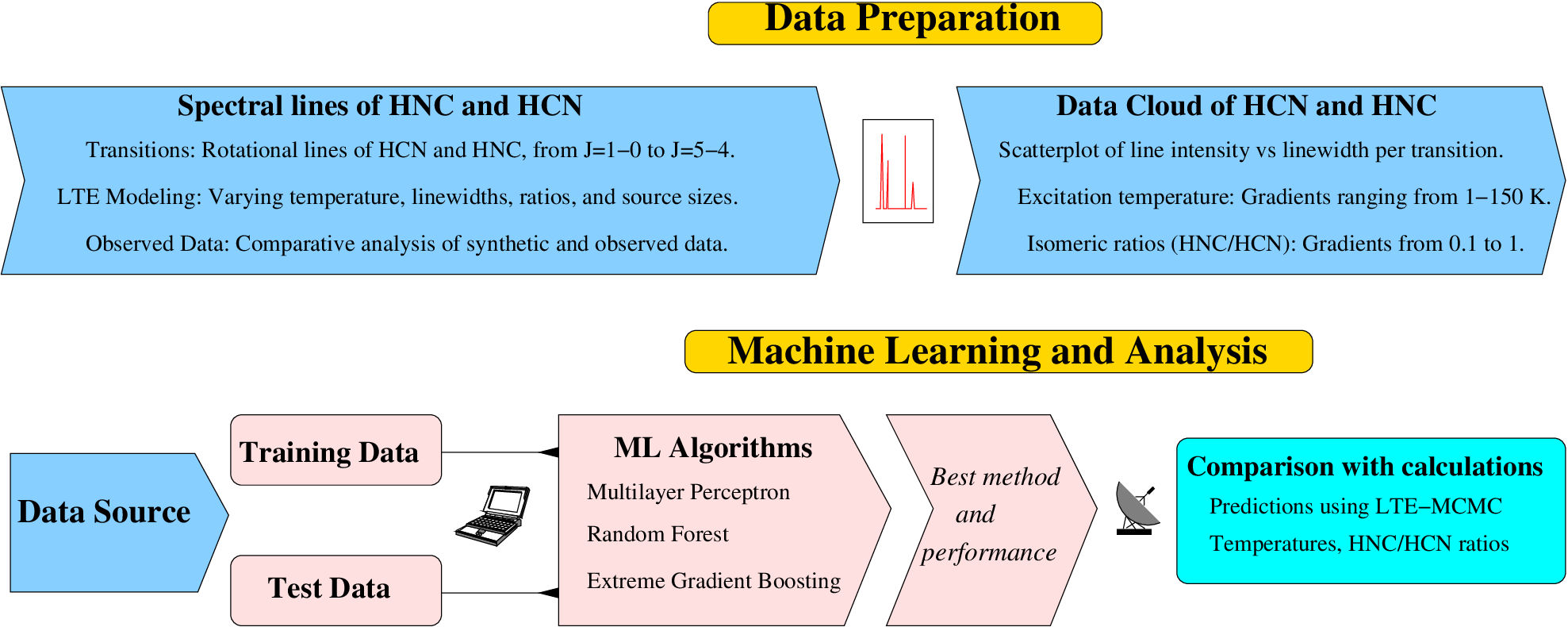}
    \caption{Schematic of the workflow for training and evaluating ML models to predict excitation temperatures and isomeric ratios from HCN and HNC spectra. Data preparation integrates models and observations to build a dataset. Gaussian fit parameters are used to generate a data cloud of line intensity vs. width, gradient-tagged by temperatures and ratios. Training and testing stages utilize three algorithms, with results benchmarked against radiative transfer models and validated using observational prototype data.}
    \label{workflow}
  \end{center}
\end{figure*}

We outlined here the methodology employed to estimate the physical conditions of an interstellar source.  Fig.~\ref{workflow} presents the workflow of the proposed bottom-up approach, which encompasses the two followed steps: data preparation and ML predictions of quantities relevant to astrophysical research. The first step, as described in Subsec.~\ref{subsec-analysis-lines}, involves the simulation of the spectral lines under different physical conditions. To do so, models were carried out under LTE hypotheses obtaining as outputs spectral lines following Gaussian distributions. A key aspect of our approach is the decision to adopt LTE. While this choice is a source of bias, it was strategically selected to establish a controlled baseline for the ML development. These simulations serve to parametrise their spectral profiles and  facilitate a comprehensive analysis of the simulated data. In the second step, ML algorithms are trained using the simulated data for estimating the physical conditions of the interstellar source.
 In this first article, particular attention is given to the HNC/HCN ratios and their excitation temperatures under LTE conditions. The
ML methods utilized in this work are outlined in Subsec.\ref{subsec-ML}. 
This methodology is developed using synthetic spectra  derived from LTE models and semiempirical data, which integrates LTE simulations with observed spectral lines. It also includes an evaluation based on observational  data collected with the APEX telescope towards the R CrA IRS 7B source, with the data reduction and analysis detailed in  Subsec.~\ref{subsec-observation-RCrA}.

\subsection{Simulation, parameterization and analysis of the spectral line profiles}
\label{subsec-analysis-lines}

We have generated an extensive set of synthetic spectra for HCN and HNC, approximately 3$\times$10$^5$ spectra, employing Gaussian line profiles as the basis for their generation. These were produced using a comprehensive grid of physical parameters pivotal for modelling spectral line profiles, including excitation temperatures, linewidths, HNC/HCN ratios, and source sizes. The compilation of the
  synthetic spectra at different ISM conditions is tackled with the parametrization of each spectral line through the intensity $I$ (K) and the  linewidth  $\Delta \nu$ (MHz). This latter is related to the linewidth at half intensity  (FWHM = $2\sqrt{2 \ln 2}\,\Delta \nu$), which is named $\Delta V_{\rm sys}$ when expressed in km s$^{-1}$.   
  This set of parameters ($I$, $\Delta \nu$) is compiled in a data cloud aimed to visualize the data distribution according to the physical conditions. The ML approach will estimate the physical conditions of a ISM source from the data distribution obtained from a thorough and comprehensive set of simulated spectral lines. 
  This is contrary to the top-down mainstream approaches (e.g., the population diagram analysis of molecular spectral lines considering the LTE approximation~\cite{goldsmith1999} or the application of the Non-LTE radiative transfer RADEX code~\citep{tak2007}), which derive the physical parameters of the observed regions from the characterization of a limited number of identified spectral lines.
In this work we propose a bottom-up approach to obtain the ISM physical conditions from the observed lines, using ML techniques and analyses based on synthetic spectra generated beforehand (see Fig.~\ref{workflow}).  Therefore, this new approach has a more general scope because, when the spectral lines simulations are carried out, it is conceived to  determine the physical parameters to a set of observed sources all at once.

The grid of the synthetic spectral lines has been generated using GILDAS~\footnote{https://www.iram.fr/IRAMFR/GILDAS} and CASSIS~\footnote{http://cassis.irap.omp.eu/?page=catalogs-vastel} software, adopting LTE conditions. Hence, the molecule energy levels are assumed populated according to the Boltzmann distribution and the excitation temperature ($T_{\rm exc}$), associated with a particular molecular transition, is equal to the gas kinetic temperature. This  can be determined from the ratio of the population or column densities between two energy levels, $i$ and $j$, characterized by their statistical weights, $g_i$ and $g_j$, and the corresponding energies, $E_i$ and $E_j$, by

\begin{equation}
\frac{N_j}{N_i}=\frac{g_j}{g_i}\exp\left[ \frac{-(E_j-E_i)}{kT_{\rm exc}}\right ].      
\end{equation}

\noindent In contrast, under non-LTE conditions, an accurate estimate of the kinetic temperature requires to consider collisional processes and to solve statistical equilibrium equations~\citep{goldsmith1999,tak2007,Shirley2015}. 

As the local properties of the gas are assumed in thermal equilibrium, a Gaussian curve is used to describe the shape of the spectral lines in the observed frequency range. This profile is associated with thermal broadening due to the random motion of molecules --specifically HCN and HNC in this case-- within the gas, leading to a characteristic velocity distribution that produces the observed spectral line shape (e.g. \citealt{roueff2021} and references therein).  Thus we can also justify the use of Gaussian fits to parametrise the simulated spectral line profiles 
and, in particular, to relate them with 
the column density of the upper level involved in the transitions and the excitation temperature. 
In this study, we have performed the simulations for the ten spectral lines $J$=1--0 (around 90~GHz), $J$=2--1 ($\sim$180~GHz), $J$=3--2 ($\sim$270~GHz) , $J$=4--3 ($\sim$360~GHz), and $J$=5--4 ($\sim$450~GHz)
of HCN and HNC at temperatures from 5~K to 150~K in steps of 5~K, abundance ratios HNC/HCN for the set of values from 0.1 to 1.0 in steps of 0.1,
source's sizes ranging from 1~arcsec to 6~arcsec with increments of 1~arcsec, and linewidths ($\Delta V_{\rm sys}$) from 1 to 19 km~s$^{-1}$ in steps of 1~km~s$^{-1}$. 
The abundance ratios HNC/HCN are computed assuming a reference value for the column density of HCN at 10$^{15}$~cm$^{-2}$, which is in accordance to the wide range of values estimated in the ISM~\citep{meijerink2011,behrens2022}.
The array of data is selected according to the surveys under study and considering the instrumental parameters of IRAM-30m~\footnote{\url{https://publicwiki.iram.es/Iram30mEfficiencies}} and  APEX~\citep{gusten2006}.
Nevertheless, other factors not considered in this work can also affect the spectral line profiles, such as the optical depth effects, beam size and resolution, source dynamics, and instrumental effects (e.g., \citealt{Kama2013,Shimajiri2015}).

In Fig.~\ref{data-distribHCN-HNC_J5-4} we show an example of how the profiles  of the spectral lines of the transitions $J=5-4$ and $J=2-1$ of HCN and HNC, encapsulated in sets of points ($I$,$\Delta \nu$), are distributed. For the sake of clarity, in this figure we have simplified the results for an abundance ratio fixed at 0.8 and a source size of 5 arcsec. Here it can be observed that the points are distributed in branches corresponding to steps in $\Delta V_{\rm sys}$: the lowest linewidth branch corresponding to $\Delta V_{\rm sys}$=1~km~s$^{-1}$ and the highest to 19 km~s$^{-1}$. 
Concerning the line frequencies and velocity conventions, rotational lines are measured in the laboratory with high accuracy. Specialized databases provide catalogues online with the rest frequencies of the spectral lines, spectroscopic parameters, and the detection status of molecules in astronomical sources\footnote{
\href{https://splatalogue.online/\#/home}{Splatalogue}, 
\href{https://cdms.astro.uni-koeln.de/}{CDMS}, 
\href{https://spec.jpl.nasa.gov/}{JPL Molecular Spectroscopy}, \href{https://physics.nist.gov/cgi-bin/micro/table5/start.pl}{NIST Recommended Rest Frequencies}}. However, the rest frequencies need to be corrected for the Doppler shifts using the observer's velocity reference frame. In this work we consider the laboratory frequencies of HCN and HNC and, to be applied to the astronomical sources, they are corrected
by the Doppler effect using the local standard of rest (LSR) of nearby sources~\citep{Binney2008}. The linewidths in velocity units stands for the broadening of these lines due to the source's motion relative to the LSR~\citep{Wilson2013}.
The sources considered in this work are embedded in the R Coronae Australis (R CrA) star-forming region, situated at $d$=149.4$\pm$0.4~pc according to data from Gaia's second release \citep{Galli2020}, and the Orion Molecular Cloud, situated at a distance of $d = 388 \pm 5$ pc \citep{Kounkel2017}.

In the upper panels of Fig.~\ref{data-distribHCN-HNC_J5-4}, the data clouds of the lines $J=5-4$ of HCN (left display) and HNC (right display) are exhibited. For these transitions, the temperature chart of the data shows us that there is a trend of the line profiles with the temperature, the larger line intensity the higher temperature. 
Nevertheless, this apparent one-to-one correspondence of the  width-amplitude diagram with the  physical quantities such as $T_{\mbox{ex}}$ is not happening for all considered transitions. In particular, the lower panels of Fig.~\ref{data-distribHCN-HNC_J5-4}, showing the profiles of the lines $J=2-1$ of HCN (left display) and HNC (right display),
exhibit a less favourable case, where
 an overlap of the data in the diagram can hinder the unravelling of the physical conditions. 
 It can be noted that the transitions of the isomers HCN and HNC involving higher excited states (i.e., $J=5-4$) have a more scattered distribution of data and, therefore, are more favourable for estimating the physical conditions at least within our adopted range. In App.~\ref{app-charts}, the  data clouds charted from  the spectral transitions $J=1-0$, $3-2$ and $4-3$ of HCN and HNC are shown (Fig.~\ref{data-distribHCN-HNC_app}).
 We conclude that it is important to know the distribution of profile points of the spectral lines 
 for estimating the survey conditions using ML.

\subsection{ML methods}
\label{subsec-ML}

The novel approach proposed in this work is thought up to be applied easily using ML techniques. ML methods are employed to determine the excitation temperatures and the HNC/HCN column density ratios independently of radiative transfer equations. They only require simple inputs derived from the Gaussian fits to the spectral lines, i.e., the line intensity and linewidth provided in units of temperature (K) and frequency (MHz), respectively.
Three different ML models, frequently used within the class of supervised problems, have been used for both classification and regression: 

\begin{itemize}
    \item Multilayer perceptron (MLP): it is the standard fully connected artificial neural network~\citep{Rumelhart1986, haykin1999}. The nonlinear activation function associated to the hidden neurons is responsible for capturing any non linear dependence in the input variables.  In this work we have first trained MLP with one hidden layer, which is supposed to be sufficient for approximating any continuous function with compact support.\footnote{This property applies for arbitrary layer width.}
    ~\citep{hornik1989}, and later on with three layers, noticing that the regression precisions slightly improved. No further improvement was noticed adding extra layers. Among the various activation functions and the optimizer algorithms available for training the network by gradient descent, the ReLU (Rectified Linear Unit) activation  function~\citep{Householder1941,Fukushima1969} and the Adam~\citep{Kingma2014} optimizer were found to yield the best performance.
    \item Random forest (RF)~\citep{ho1995,Breiman2001}: it belongs to the class of Ensemble algorithms, where the predictions of several Decision Trees are averaged as the final outcome. The different estimators are trained by randomly selecting subsets of the input features and also by changing the training set via bootstrap aggregating, a random subsampling with replacement that helps avoiding overfitting and reducing variance. 
    \item Extreme gradient boosting (XGB)~\citep{Schapire2003, chen2016}: it is a very efficient, open source implementation of the ensemble gradient boosting type of algorithm, in which weak learners, e.g. decision trees with a single split, are recursively added in order to improve the predictions made by the previous learners. The final outcome is an averaged result, weighted by the individual accuracy of the weak learners. The weights are adjusted by minimizing the loss function, which means that this model includes a learning rate parameter in the same fashion as it is included in the gradient descent algorithms used to fit the weights of a MLP.
\end{itemize}

 These three models have been trained using 80~\% of the simulated spectral lines as training sample obtained by varying the physical conditions of the input model. The remaining 20\% of the data was used to test the predictive power of the trained model. Each model was trained 100 times from scratch selecting randomly a different sampling partition each time. This procedure intended to minimize the bias introduced by the choice of a particular data splitting. This Monte Carlo cross-validation (see, e.g., \citealt{Burman1989, XU2001}) with 100 iterations has also been used to optimize the hyperparameters of the models. We stress that for each combination of input spectral lines a separate model, with its own hyperparameters, is trained using only that particular combination of synthetic lines as training dataset. Since the features of each line have been encoded in its two Gaussian parameters (linewidth and intensity), a model trained to make predictions from $n$ spectral lines takes as input a $2n$-dimensional vector containing the Gaussian parameters stacked together.
 These input vectors have been rescaled, before being fed to the ML models, using the standard scaler that transforms the input features by gauging the training data to have zero means and unit standard deviations~\citep{Sola1997,Jain2000,Pedregosa2011}. 
 
 Although a vanilla cross-validation approach for tuning the hyperparameters can still  introduce some bias, we did not see the need for introducing a further splitting into a validation and testing datasets, nor for implementing a more computationally demanding nested cross-validation approach, for the following reasons: the distribution of the training Gaussian parameters and corresponding target values are rather uniform and noiseless (see Figs.~\ref{data-distribHCN-HNC_J5-4}, \ref{data-distribHNC_J4-3} and \ref{data-distribHCN-HNC_app}) since they correspond to Gaussian fits of synthetic lines generated by uniformly varying the target values. If it turned out that some residual bias had not been washed out by the Monte Carlo cross-validation, this should be negligible compared to the bias introduced by optimistically assuming that the noisy data corresponding to real spectral lines will follow the same distribution. An estimate of this bias goes beyond the scope of this first exploratory analysis, aimed at testing the soundness of the ML predictions on a few real spectral lines.  
 Finally, we also stress that the hyperparameters optimization has been carried out by maximizing the average of the precision metric evaluated on the training data subsets, but with the constraint that the average of the metric evaluated on the testing data subsets does not deviate considerably, therefore avoiding overfitting. 

 On the one hand, the relevant hyperparameters for the MLP models have been identified in the number of layers, the number of neurons per layer, the learning rate of the optimization algorithm and the $\alpha$ coefficient of the L2 regularization implemented to prevent  overfitting~\citep{Girosi1995}. The best choice, following the criteria commented in the previous paragraph, was found using three layers with 32, 64 and 128 neurons respectively, a learning rate of $10^{-3}$ (the default value for the Adam algorithm, see, e.g.,\citealt{Diederik2014}) and $\alpha=10^{-4}$. These values were found to be robust in yielding the best results across the different combinations of input spectral lines.

 On the other hand, it was found that the optimal relevant hyperparameters for RF and XGB needed to be adjusted for each separate training corresponding to different input lines, with the exception of the number of estimators which has been fixed to 500 throughout every training for both RF and XGB. The only additional relevant parameter for RF was found to be the depth of the forest, while for XGB was found to be the combination of the depth and the learning rate.
 Since the values of these hyperparameters can vary according to each particular combination of input lines, we will show their values for each specific prediction. 

 Once the hyperparameters have been optimized, the predictions for the transitions of the real spectral lines have also been averaged over 100 independent trainings, evaluating their statistical standard deviations in order to assess the degree of reproducibility of the predictions as well as use them as uncertainties of the target quantities.
 In the regression fit, the models are trained by minimizing 
 the standard Mean Squared Error (MSE) on the training dataset:
 
\begin{equation}
\mathrm{MSE} = \frac{1}{N} \sum_{i=1}^N (y_i - \hat y_i)^2,
\end{equation}

\noindent where N is the size of the training sample, $y_i$ and $\hat y_i$ are the label target values and the predicted target values, respectively.
As a metric to assess the performance of the model, we used the coefficient of determination, $R^2$:

\begin{equation}
\label{coeffdet}
R^2 = 1 - \frac{\sum_{i=1}^N (y_i - \hat y_i)^2}{\sum_{i=1}^N (y_i - \bar y)^2} ~~,
\end{equation}

\noindent where $\bar y$ is the mean value of the true target values.

\subsection{Evaluation with R CrA IRS 7B observations}
\label{subsec-observation-RCrA}

The evaluation of this new approach is carried out using APEX data taken from the ESO Science Archive Facility~\citep{gusten2006,wampfler2014}. In particular, this work has been focused on Class 0 young stellar object R CrA IRS7B (RA=19\(^h\) 01\(^m\) 56.\(^s\)4, DEC=-36$^\circ$ 57' 28.3''), which was observed with APEX between August 16, 2012, and October 1, 2012. We selected this source because the two isomers HCN and HNC have been detected in there and few of their transition lines were identified. 
In fact, the $J$=3--2 rotational transitions of HCN, H$^{13}$CN, HC$^{15}$N, HNC, HN$^{13}$C, and H$^{15}$NC have been observed. It makes the source R CrA IRS7B a good candidate to assess  our new approach.

The data archive for this study was collected using the Swedish Heterodyne Facility Instrument (SHeFI) single-sideband SIS receiver APEX-1 (211--275 GHz), as described by \citep{gusten2006,vassilev2008}. This receiver was combined with the eXtended bandwidth Fast Fourier Transform Spectrometer (XFFTS). The broad bandwidth of this backend allows for simultaneous observation of several isotopologues, minimizing calibration uncertainty based on relative rather than absolute calibration across the band. This study utilized archival observations of R CrA IRS 7B, conducted in August 2012. In addition, the absolute calibration uncertainty was taken into account with a  margin of up to 30~\%, according to the 
issues related to the H$^{15}$NC (3--2) observations, for which an equal calibration uncertainty was reported~\citep{wampfler2014}.

The spectra provided by APEX are formatted in GILDAS/CLASS using the corrected antenna temperature scale ($T_{\rm A}^{\ast}$). To convert these intensities to the main beam temperature scale ($T_{\rm MB}$), the spectral setups were calibrated using a forward efficiency of 0.95 and a beam efficiency of 0.75 for APEX-1.
Each data set was examined for spectral anomalies in its observations to avoid potential issues affecting the HCN and HNC lines. Within the spectral windows, intense lines of HCO$^+$ (3--2) were also identified. Therefore, the spectral profiles of HCN, HNC, and HCO$^+$ were reviewed within the observation set prior to the baseline extraction. First-order polynomial baselines were applied to the individual scans. Subsequently, all scans were averaged using weights equal to 1/$\sigma_{\rm tn}^2$, where $\sigma_{\rm tn}$ is the standard deviation of the thermal noise in the data. This weighting performed on GILDAS-CLASS assigns higher weight to data points with lower uncertainty, minimizing the impact of noisier measurements and resulting in a more accurate and reliable average spectrum. After the baseline correction of the  data, the spectra of HCN and HNC were identified and adjusted  using Gaussian functions. This process was used to estimate the integrated line intensity ($W$ in K km s$^{-1}$) along with its associated uncertainty. However,  to compute the overall error ($\Delta W$), it is necessary to account for  the calibration uncertainty, as expressed by the following formula:

\begin{equation}
\Delta W = \sqrt{(cal/100 \times W)^2 + (rms \sqrt{2 \times \Delta V_{\rm sys} \times \Delta v})^2},    
\end{equation}

\noindent where $cal$ is the calibration uncertainty (\%), $rms$ is the noise around the line, $\Delta V_{\rm sys}$ is equivalent to the FWHM (km s$^{-1}$) and $\Delta v$ is the bin size (km s$^{-1}$).
Estimates and uncertainties of parameters, such as column densities and excitation temperatures,  are obtained by incorporating both types of errors using the Line Analysis module and scripts within the CASSIS software \citep{wampfler2014,Vastel2015}.

\section{Results}
\label{sec-results}

The analysis of LTE spectral line profiles of HCN and HNC relative to the physical conditions (Subsec.~\ref{subsec-analysis-lines}) has allowed us to take a step forward in the process of ML estimate of the physical conditions. First of all, we have assessed the soundness of the ML algorithms using semiempirical spectra from Orion KL hot core (Subsect.~\ref{subsec-finetuning}). This trial-and-error approach has allowed us to carry out subsequently the estimates from the observed spectral lines of R CrA IRS 7B region (Subsect.~\ref{subsec-RCrA}).

\subsection{ ML predictions using semiempirical data from Orion KL hot core}
\label{subsec-finetuning}

Following  the preliminary analysis of the line profiles  described  in Subsec.~\ref{subsec-analysis-lines}, the  subsequent step is to apply this ML-based approach to determine the temperature and HNC/HCN ratio  of the interstellar source. 
In order to validate this new approach, it has been applied to semiempirical APEX spectra of the Orion KL hot core. These spectra were derived by combining observational data, incorporating spectral noise, with synthetic LTE models to resemble the heated gas components in the Orion KL hot core~\citep{wright2017}. Thus, the real data are used as a template to guide the LTE computation within the hot core excitation conditions, ensuring that the noise and systemic velocity characteristics of the observed data are accurately preserved. In Fig.~\ref{fig:hcn-hnc-ori} we show the semiempirical spectral lines of HCN and HNC used in this work. These spectral lines have been modelled using Gaussian functions to simplify the representation, reflecting an idealized case. However spectral lines are often influenced by various factors such as physical effects, e.g., Doppler and collisional broadening, shocked gas or molecular outflows \citep{mendoza2018,Hervias2019,Guerra-Varas2023}, and instrumental/data processing issues, e.g., finite resolution,  spurious spikes, or inadequate baseline subtraction \citep{dumke2010,Stanke2022}. Given these factors, Gaussian functions may not always suffice, and in certain cases, Lorentzian or Voigt functions may provide a more accurate fit, particularly when both Doppler and collisional broadening are present, while kurtosis-based distributions may be needed to account for asymmetries or complex profiles (e.g. \citealt{Juvela2024}). A detailed examination of these modelling functions and their applicability will be addressed in future studies, as our method is further developed.

The Orion KL nebula is located in the Orion Molecular Cloud-1 (OMC-1) \citep{Kounkel2017}, which is a key prototype source for studying high-mass star formation, featuring infrared cores, H$_2$O masers, millimetre continuum emission, compact radio sources, molecular outflows and hot cores \citep{Dalei2020}. Additionally, it is significant for astrochemical studies due to its complex molecular composition and the physical-chemical conditions present in its lukewarm heated gas components~\citep{blake1987,schilke1997,comito2005,Brouillet2015}. 
In this system, different molecular abundances have been associated with gas components at different positions. \citet{Taniguchi2024} estimated low and high DCN/DCO$^+$ ratios in gas components at $V_{lsr} \approx 7.5$–$8.7$ km s$^{-1}$ and $V_{lsr} \approx 9.2$–$11.6$ km s$^{-1}$, respectively. Considering an \ \lq\lq abundant\rq\rq \ region in both HCN and HNC, we have adopted a systemic velocity of $V_{lsr} \approx 9.4$ km s$^{-1}$ for the data used in the validation of this new approach.

Three semiempirical spectra of Orion KL hot core, HCN $J$=5--4, HNC $J$=3--2, and HNC $J$=4--3,  have been analysed for this subsection and are displayed in Fig.~\ref{fig:hcn-hnc-ori}. Given that these lines consist of calibrated APEX data alongside predictions from LTE models, we have considered a calibration uncertainty of up to 19~\% during the line analysis and routines, specifically for the frequency range of the transitions examined here. In fact, the assumed calibration uncertainty is relatively high for the purpose of this work. However, this is consistent with the literature, where lines of CO, HCN, H$_2$CO and CH$_3$OH  can exhibit significant  calibration uncertainties across various frequencies \citep{dumke2010}. The transition lines obtained with the LTE model are also exhibited in Fig.~\ref{fig:hcn-hnc-ori}, which provided the conditions of $T_{\rm exc} =$ 90~K and $N$(HNC)/$N$(HCN)=~0.8. These spectral lines were considered with a rms of $\lesssim$~800~mK. This noise level refers to the spectral channels around the selected transitions of HCN and HNC. The channels, each of 0.03~km~s$^{-1}$ width, span spectral windows ranging from $\sim$~-5 to 25~km~s$^{-1}$, centred around the systemic velocity of Orion KL ($V_{lsr}\thickapprox$ 9.4~km~s$^{-1}$).

Here ML techniques are applied to determine the physical conditions of Orion KL hot core from the data cloud obtained from the spectral lines profiles. While ML approaches have been previously employed in the fields of molecular astronomy and astrochemistry, they have been used for different purposes, e.g., to create interstellar chemical inventories \citep{lee2021}, predict ionization fraction \citep{Bron2021}, predict the intensity of the incident UV field \citep{Gratier2021}, and apply denoising methods to low signal-to-noise ratio data cubes \citep{einig2023}. However, these references represent just a small selection of the work done within the ISM community. 
 In this paper, the data cloud generated and collected as  sets of ($I$,$\Delta \nu$) from the spectral line profiles simulated under LTE conditions are taken as the input features for  the ML approaches.
The target variables~\footnote{In Machine Learning, the target variables are the output variables.} are the excitation temperature ($T_{\rm ex}$) and the ratio HNC/HCN, i.e. the quantities that the models will be able to predict once properly trained. Since these are continuous variables, we set out the ML inference task as a regression problem. 

The results of the physical conditions for Orion KL hot core are derived from the analysis of the data cloud and the predictions obtained with the regression performed using ML methods. The training
of the ML  models has been carried out using the
aforementioned grid of values obtained from Gaussian fits (line intensities and linewidths). Once trained, the models can make predictions on the observed spectra,  taking as input their Gaussian parameters and providing the values of the target variables of the source (excitation temperature and the HNC/HCN ratio).

In Fig.~\ref{data-distribHNC_J4-3}, the distribution of data is exhibited for the profiles of the simulated spectral lines of HNC $J$=4--3 obtained for the grid of physical parameters' values. 
We present here the data chart for the HNC $J$=4--3 transition to emphasize the importance of the data cloud distribution for accurately predicting the HNC/HCN isomeric ratios.
On the  top display of Fig.~\ref{data-distribHNC_J4-3}, the temperature mapping of the data cloud is given. On the bottom one, the data are charted with the abundance ratio. 
In addition, in both panels of Fig.~\ref{data-distribHNC_J4-3}, a green dot representing the line intensity and linewidth values,  with their corresponding error bars,  has been included along with the  HNC (4--3) line of Orion KL.  This illustrates the placement of their parameter values ($I, \Delta \nu$) within the data chart used to determine the physical conditions during the ML optimization process. These values have been derived 
from the Gaussian fit to the observed spectrum 
adopting a total error estimate which is dominated, in this case, by a high calibration uncertainty (19~\%). For this observed line, the  intensity is 70 $\pm$ 13 K and the linewidth is reported as 2.2 $\pm$ 0.4 km s$^{-1}$ in velocity, and 2.6 $\pm$ 0.5 MHz in frequency. 

As it can be observed in Fig.~\ref{data-distribHNC_J4-3}, the mapping of the simulated spectral lines' profiles into the data cloud given by $I$  and $\Delta \nu$ becomes cumbersome to visualize when the number of physical parameters considered in the simulations increases. 
Thus, to simplify the chart of the data cloud and facilitate the inputs into the ML computations, estimations were obtained for each source size value, with a particular emphasis on a source size of 5~arcsec. 
In earlier studies, \citet{Vicente2002} have discussed the physical conditions of the heated gas in the Orion KL hot core, considering a source size value of 5~arcsec. In comparison with Fig.~\ref{data-distribHCN-HNC_J5-4}, the branches of Fig.~\ref{data-distribHNC_J4-3} are broadened because of the variation of the abundance ratio.

A higher number of identified spectral lines can lead to improve the $R^2$ precision and the parameters robustness when all lines contribute comparably to constraining target variables. An optimal line selection --prioritizing the strategic tracer/transition combination over the mere quantity-- can determine the physical parameters more efficiently while preserving computational feasibility, as \citet{einig2024} discussed. For the semiempirical HCN/HNC spectra of Orion KL hot core, we should consider the synthetic data for the three identified transitions (HCN $J$=5--4; HNC $J$=3--2 and 4--3).
In fact, it can also be useful to consider different combinations of these three transitions to assess the estimates of the physical quantities as well as their precisions. This will increase the dimensionality of the features space and can contribute to improving the accuracy of the prediction, even if the distribution of the lines show a strong overlap. Nevertheless, if only a single line
is identified in the source, we should not use it as an input feature whenever its data distribution has a strong degree of overlap, according to the preliminary analysis of the data cloud. 
This approach for estimating the temperature and abundance ratios in the ISM can be significant for those surveys where only one spectral line is identified~\citep{roueff2021}, subject to a comprehensive analysis of the spectral lines' profiles.

The three  Machine Learning models MLP, RF and XGB have been trained. Each model was evaluated over the independent 100 training samples, therefore, the mean values of the coefficient of determination, $\bar R^2$, corresponding to the training and testing sets are compared for the three different models. During the assessment of  the predictions of the semiempirical lines, we realized that, for MLP and XGB models, some combinations of transitions provide unphysical negative values for the temperature
and/or the abundance ratio. This is caused because, first, MLP and XGB models can extrapolate and provide results out of the range of training data values, and second, the set ($I$,$\Delta \nu$) of the observed lines, in case of multiple lines are considered whose Gaussian parameters are stacked into a multidimensional vector, are close to the border of the data distribution and out of the convex hull~\footnote{The convex hull defined by a set of vector points in any dimension is the smallest convex set that contains them. It is a geometrical construction (not the only one) used to determine whether a point interpolates or not the training dataset domain \citep{Brooks01081988}.} of the set of training data. We have overcome this issue  by fitting the models to the natural logarithms of the target values and, afterwards, transforming the predictions to the original scale by taking their exponential. We stress however that the conversion back to the natural scale values has been performed only for showing the predicted values of the real (or semiempirical) spectra. During every other step, including the evaluation of the $R^2$ values, the logarithmic scale of the target values has been maintained  in all the training/testing phases.
 In Tables~\ref{tab_temp-ML} and \ref{tab_abundance-ML}
the comparison among the three models RF, XGB and MLP is carried out. 
We show that the results of the three models are  in general alike and, therefore, it makes us expect that this novel approach qualifies for predicting the excitation temperature and the abundance ratio of a source. Nevertheless, in this case, we  will overlook the results of MLP model because we checked that it tends to extrapolate to values out of the set of training data. Thus, we will mainly focus on the results of RF and XGB, which both were proven equally suitable.

 The results of the excitation temperature as target variable for these three ML models are given in Tab.~\ref{tab_temp-ML} as well as the training and testing values obtained for the coefficient of determination $\bar R^2$.
It can be observed that, based on the values of the coefficients of determination, the results for the three ML models obtained only with the spectral line HCN $J$=5--4 are much more reliable than  considering the line HNC $J$=4--3. In this case, none of the models are useful when
the single transition $J$=4--3 of HNC is considered because their coefficients of determination are well below. The low coefficient of determination for HNC $J$=4--3 stems from the overlap of the distribution of the spectral line profiles (see  top display in Fig.\ref{data-distribHNC_J4-3}).  Therefore, the preliminary analysis of the spectral distribution is useful to select, before the ML optimization process, those spectral lines  that do not have a strong superposition of the branches in the data cloud distribution. Nevertheless, when we combine the two lines involving both isomers (HCN $J$=5--4 and HNC $J$=4--3), the result in Tab.~\ref{tab_temp-ML} improves slightly with respect to the one obtained for HCN $J$=5--4, in compliance with the mean values of the coefficients of determination. This combination of two lines is actually implemented by stacking their Gaussian parameters in a four-dimensional input vector. This increase of the features space dimensionality permits to get rid of the ambiguity caused by the overlap present in a two-dimensional space and therefore to improve the performance.

When a single transition $J$=5--4 of HCN is considered, the three models provide similar results of the excitation temperature (see Tab.~\ref{tab_temp-ML}) but the RF and XGB models are the most suitable concerning the somewhat higher values of $\bar R^2_{train}$ and $\bar R^2_{test}$. The temperature predictions of RF, XGB and MLP for $J$=5--4 of HCN are $90\pm0$~K, $90\pm1$~K and $93\pm3$~K, respectively.
When the two transitions $J$=5--4 of HCN and $J$=4--3 of HNC are used, the predictions of RF and XGB are almost identical, $90\pm1$~K and $89\pm1$~K
(with coefficients of determination $\bar R^2_{train}$=0.999 and ${\bar R}^2_{test}$=0.998), whereas the prediction of MLP deviates from the expected value. 
It seems that the extrapolation that MLP model makes is more pronounced than for the other two models and, when we use the two lines $J$=5--4 of HCN and $J$=4--3 of HNC, the real data are pinpointed outside the convex hull defined by the set of training data, affecting MLP more than the other two models. 

 Concerning the  HNC/HCN ratio, the predictions and the values of $\bar R^2_{train}$ and $\bar R^2_{test}$  are given for the three ML models in Table~\ref{tab_abundance-ML}. The same three ML models are trained using the ratio HNC/HCN as a target variable. 
In Table~\ref{tab_abundance-ML}, the HNC/HCN ratios derived using the three ML models are based exclusively on HNC lines, specifically the lines HNC $J$=3--2, $J$=4--3 and a combination of these two transitions. The HCN $J$=5--4 line was not included in these calculations. In this case, a constant HCN column density is assumed and, thus, the estimated ratio primarily reflects the variations in the HNC column density.
It can be noted in Table \ref{tab_abundance-ML} that the results are similar for the three approaches although the one involving the combination of lines HNC $J$=3--2 and HNC $J$=4--3 improves the result of HNC/HCN relative to the values of the coefficients of determination.

 When a single transition line is taken into consideration, the results of the HNC/HCN ratio are alike for the three models, obtaining for HNC $J$=3--2 an abundance ratio of $0.83\pm0.02$, $0.82\pm0.02$ and $0.82\pm0.04$, with $\bar R^2_{train}$=0.77, 0.81 and 0.77 and ${\bar R}^2_{train} - {\bar R}^2_{test}$=0.03, 0.02 and 0.02 for the models RF, XGB and MLP models, respectively. The results of the transition line $J$=4--3 of HNC are also close to those obtained for $J$=3--2 but with smaller values for the coefficients of determination. When the two transitions $J$=3--2 and $J$=4--3 of HNC are taken, 
the models RF and XGB also provide close values for the abundance ratio, 0.81 and 0.82, with $\bar R^2_{train}$= 0.86 and 0.96 and ${\bar R}^2_{test}$=0.84 and 0.94  for both models, respectively. As for the estimate of $T_{\rm exc}$, the MLP result for the HNC/HCN ratio is 
also smaller. In this case, the ratio is 0.72, with $\bar R^2_{train}$ and ${\bar R}^2_{test}$ equal to 0.96, and it will be again overlooked because of its extrapolation tendency out of the set of training data. 

Next we make a comparison between the results obtained with this novel approach and with the LTE model.
Semiempirical data from Orion KL have been utilized here to  benchmark three different ML models. 
On the one hand, the three lines --HCN $J$=5--4, HNC $J$=4--3, and $J$=3--2-- have been modelled  simulating the excitation condition of Orion KL hot core (Fig.~\ref{fig:hcn-hnc-ori}) under LTE conditions providing a rough solution for the temperature and HNC-to-HCN ratio of approximately 90~K and 0.8, respectively. 
Consequently, these LTE results primarily serve to evaluate the performance and accuracy of the ML models, allowing for an initial validation of their predictive capabilities under these specific astrophysical conditions.
On the other hand,  according to the results exhibited in  Tables~\ref{tab_temp-ML} and \ref{tab_abundance-ML}, we can state that the RF and XGB algorithms are equally good for this case and, therefore, they provide the most reliable ML predictions. When two transition lines are considered, RF and XGB 
have accomplished an excitation temperature  of $90\pm1$~K and $89\pm1$~K and an isomeric ratio of $0.81\pm0.02$ and $0.82\pm0.02$, respectively. Therefore, the ML models, trained and tested on both synthetic and semiempirical datasets, have demonstrated their capability to replicate the physical conditions using only a few inputs. However, the success of these predictions is contingent upon whether the data cloud generated by parametrizing the LTE spectra (Fig.~\ref{data-distribHNC_J4-3}) is well-defined.

In the literature, sub-mm 
observations of the Orion hot core provided estimates of HNC/HCN ratios around 0.01 and excitation temperatures above 100~K~\citep{Nickerson2021}. Towards the Integral Shape Filament in Orion, a correlation between the intensity ratios of HNC/HCN, ranging from $\sim$~1 to 0.07, and gas kinetic temperatures derived from NH$_3$, ranging from $\sim$~10 to 90~K, has been found~\citep{hacar2020}. In OMC-1, the HNC/HCN ratios might vary  between 1 and 0.01. Although the abundances of HNC in OMC-1 are similar to those in dark cloud cores, its abundance is notably lower in regions characterized by higher temperatures \citep{schilke1992}. Surveys conducted towards dark cloud cores have unveiled a wide range of HNC/HCN ratios, with reported values ranging from 4.5 in L1498 to 0.54 in L1521E \citep{hirota1998}. In addition, according to the observations and chemical modeling,  HNC-to-HCN line intensity ratios of up to 0.8 have been detected in the outer regions of protoplanetary disks~\citep{Long2021}, akin to the HNC/HCN ratios explored in this study. In external galaxies, \cite{Perez2007} conducted observations of the HCN and HNC $J$=3-2 lines in a sample of luminous Seyfert galaxies with prominent HNC emission. Although their results generally indicate higher abundances of HCN, they identified a source where the HNC/HCN J=3-2 line ratio was larger than unity. In the case of the ultraluminous infrared galaxy IRAS 20551–4250, \citet{Imanishi2017} found, using ALMA data, a particular result concerning the HNC emission. They observed that higher rotational excitation of HNC compared to HCN and HCO$^+$ is difficult to explain without taking into account a scenario  involving infrared radiative pumping. In this study, the ML predictions for the Orion KL hot core show a reasonable agreement with the results from the literature, despite differences in observational setups and radiative scenarios. To demonstrate the robustness of our proposed method in a realistic context, we analyse the APEX observations of the source R CrA IRS 7B, presenting the results and comparing them with estimates derived from the LTE approximation, as discussed in the further section.

\subsection{Application of the model to the source R CrA IRS 7B}
\label{subsec-RCrA}

The R CrA region is among the closest and most dynamic star-forming regions in the Solar neighbourhood. Recent surveys have identified 393 young stellar object candidates, which are relatively evolved and classified as Class II and III sources \citep{Esplin2022}. The analysis presented here focuses on line observations of the R CrA IRS 7B source. However, it is worth noting that there are neighbouring sources located at a short spatial separation~\citep{Schoier2006}.
More recent studies based on high-resolution maps and images have revealed the presence of such a protobinary system and the presence of stellar companions in the direction of R CrA \citep{Yang2018,Mesa2019,Perotti2023}.  

Concerning the R CrA IRS 7B source, we have conducted the treatment and analysis of the observed spectral data of the $J$=3--2 transitions from the isotopologues of HCN and HNC, which were retrieved from the ESO Archive~\citep{wampfler2014}.
In this work, we have used these observations to estimate the excitation temperatures and HNC-to-HCN ratios. 
For the purpose of comparing the results of the ML-based approach with a mainstream one, we have calculated them under the LTE assumption. This latter approximation
is justified by the fact that the observed spectral lines are partially optically thin and generally conform to LTE conditions  (see \citealt{wampfler2014}). These calculations are based on analyses of the $J$=3--2 transitions of HCN, HNC, H$^{13}$CN, HN$^{13}$C, HC$^{15}$N, and H$^{15}$NC. In Subsec.~\ref{subsubsec-LTEcalculations} we provide the LTE results for R CrA IRS 7B. In Subsec.~\ref{subsubsubsec-ML-RCrAIRS7B}, we present the results obtained with ML for HCN and HNC spectra and make the comparison with the LTE approximation.

\subsubsection{LTE calculations of HCN and HNC in R CrA IRS 7B}
\label{subsubsec-LTEcalculations}

The transition lines $J$=3--2 of the main species and $^{13}$C and $^{15}$N isotopologues of HCN and HNC have been identified for R CrA IRS 7B within an APEX spectral setup ranging between $\sim$~258 and 272~GHz. LTE calculations were conducted using the Markov chain Monte Carlo (MCMC) algorithm, a robust method for sampling from probability distributions \citep{Foreman2013}. This approach allows for the numerical estimation of excitation temperatures and a crucial parameter in astrophysical models such as molecular column densities. 
By treating these parameters as free variables, the LTE-MCMC algorithm employs chi-squared ($\chi^2$) minimization to fit the model to observational data accurately. As a result, the calculations not only yield accurate estimates of $N_{\rm col}$ and $T_{\rm rot}$, summarised in Table~\ref{tab:lineanalysis}, but also produce a model that fits the observed spectral lines shown in Fig.~\ref{fig:hcn-hnc-rcr}. 

In Table~\ref{tab:lineanalysis}, the corresponding spectroscopic and modelled parameters of the spectral lines of HCN and HNC isotopologues observed in R CrA IRS 7B  are outlined. The 
upper energy levels corresponding to the analysed lines span from  24 to 27~K, with Einstein coefficients ranging approximately from 6.89 to 9.34 $\times$ 10$^{-4}$ s$^{-1}$. 
Table~\ref{tab:lineanalysis} also presents the results of the LTE-MCMC analysis. HCN exhibits the highest column density at 3.9 $\times 10^{13}$ cm$^{-2}$ and an excitation temperature of 26~K. HNC, with a slightly higher excitation temperature of 27~K, has a lower column density of 1.17 $\times 10^{13}$ cm$^{-2}$. 
The excitation temperatures for the other isotopologues range from 11 K to 14 K and show progressively
decreasing column densities, with H$^{13}$CN at 4.0 $\times 10^{12}$ cm$^{-2}$,  HN$^{13}$C at 1.5 $\times 10^{12}$ cm$^{-2}$, HC$^{15}$N at 1.0 $\times 10^{12}$ cm$^{-2}$, and H$^{15}$NC at 3.9 $\times 10^{11}$ cm$^{-2}$.
From the ratios of the column densities $N_{\rm col}$ of HNC and HCN isotopologues, the  HNC/HCN turns out to be $0.30\pm0.03$ for HNC/HCN, $0.38\pm0.08$ for HN$^{13}$C/H$^{13}$CN and $0.39\pm0.25$ for H$^{15}$NC/HC$^{15}$N. 
Regarding the ratios between the isomers, it is evident that HCN is around three times more abundant than HNC. In terms of isotopic fractionation, the most abundant isotopologues are those with $^{12}$C, followed by $^{13}$C and $^{15}$N, aligning with expected trends~\citep{wilson1994}. 

\begin{table*}[h!]
\centering
\caption{Spectroscopic parameters of the observed lines towards  R CrA IRS 7B, including Gaussian-integrated areas and LTE-MCMC derived results for HCN and HNC and their $^{13}$C and $^{15}$N isotopologues.}
\begin{tabular}{lccccccc}
Species & Frequency (MHz) &  $E_{\rm up}$ (K) & $A_{\rm ij}$ (10$^{-4}$ s$^{-1}$) & $\int T_{\rm mb}dv $ (K km s$^{-1}$)& $T_{\rm exc}$ (K) & $N_{\rm col}$ (cm$^{-2}$) &  $\tau$ \\
\hline
HCN	(3--2)	&	265886.434	&	25.52	&	8.36	&	29.4	(2)	&	26	(5)	&	3.9	(3)	$\times$	10$^{13}$	&  0.60\\																			HNC	(3--2)	&	271981.142	&	26.11	&	9.34	&	12.0	(1)	&	27	(5)	&	1.17	(9)	$\times$	10$^{13}$	&  0.30\\													
H$^{13}$CN	(3--2)	&	259011.798	&	24.86	&	7.73	&	1.85	(3)	&	11	(3)	&	4.0	(2)	$\times$	10$^{12}$	&  0.20\\															
HN$^{13}$C	(3--2)	&	261263.513	&	25.08	&	8.28	&	0.80	(3)	&	12	(2)	&	1.5	(3)	$\times$	10$^{12}$	&  0.10\\
HC$^{15}$N	(3--2)	&	258156.996	&	24.78	&	7.65	&	0.45	(2)	&	11	(4)	&	1.0	(6)	$\times$	10$^{12}$	&  0.06\\
H$^{15}$NC	(3--2)	&	266587.800	&	25.59	&	6.89	&	0.21	(3)	&	14	(2)	&	3.9	(8)	$\times$	10$^{11}$	& 0.03\\
\hline
\end{tabular}
\label{tab:lineanalysis}
\tablefoot{Numbers in parentheses represent the uncertainties.}
\end{table*}

In comparison with the literature, \cite{watanabe2012} conducted an LTE analysis of spectral lines, including HCN and HNC, towards the same source, using data from the ASTE 10m telescope. They computed their column densities assuming values of the excitation temperatures comparable to our results, of 15 K, 20 K, and 25 K.
In particular, at 25~K, Table 3 of \cite{watanabe2012} reports column densities for HCN and HNC of 7.1 $\times 10^{13}$ cm$^{-2}$ and 2.3 $\times 10^{13}$ cm$^{-2}$, and HNC-to-HCN ratios derived from these column densities of $0.33\pm0.11$ for 15~K and $0.32\pm0.11$ for 20~K and 25~K. These results are in general consistent with our estimates. 
In addition, the excitation temperatures observed here are consistent with those typical of a relatively cold environment. For comparison, \cite{Schoier2007} estimated gas temperatures of CH$_3$OH and H$_2$CO around 20 K and between 40–60 K, respectively. In the case of \cite{watanabe2012}, their study reported temperatures ranging from $\sim$~16 K, for CCH, to 31 K, for CH$_3$OH.

The lines of the most abundant isotopologues often do not satisfy the optically thin approximation \citep{mangum2015}. For the purpose of this work, which is based on LTE assumptions, the $J$=3--2 spectral lines of HCN and HNC in R CrA IRS 7B are reasonably well modelled under LTE conditions.
Fig.~\ref{fig:hcn-hnc-rcr} is presented with the observed and modelled spectra, with intensities shown in their default units of antenna temperature. The line intensities and optical depths of these transitions of HCN and HNC are found to be approximately proportional to the column density, with calculated optical depths ($\tau$) of $\sim$~0.6 and 0.3, respectively. The scenario is similar for the less abundant isotopologues. For the $J$=3--2 transitions of H$^{13}$CN and HN$^{13}$C, the values were $\tau \thickapprox$ 0.2 and 0.1, respectively. Likewise, for $J$=3--2 transitions of HC$^{15}$N and H$^{15}$NC, $\tau \thickapprox$ 0.06 and 0.03, respectively.

In addition, an inspection of nitrogen isotopic fractionation has been carried out using the double isotope method. This analysis was conducted under the assumption that all lines are optically thin and adopting a $^{12}$C/$^{13}$C ratio of 69 according to the local ISM standards~\citep{Milam2005,wampfler2014}. The $^{14}$N/$^{15}$N ratios have been calculated from the ratio of the integrated lines of two singly substituted isotopologues of HCN or HNC, multiplied by the $^{12}$C/$^{13}$C ratio for the local ISM. Based on the integrated areas listed in Table~\ref{tab:lineanalysis}, the $^{14}$N/$^{15}$N ratios are about 284 for H$^{13}$CN(3–2)/HC$^{15}$N(3–2), and about 262 for HN$^{13}$C(3–2)/H$^{15}$NC(3–2). These results agree with those reported by \cite{wampfler2014}, who obtained H$^{13}$CN(3--2)/HC$^{15}$N(3--2) $\thickapprox$ 287 and HN$^{13}$C(3--2)/H$^{15}$NC(3--2) $\thickapprox$ 259, as well as H$^{13}$CN(4–3)/HC$^{15}$N(4–3) $\thickapprox$ 285 for a different transition (e.g. see also \citealt{watanabe2012}).

\subsubsection{Results of ML approaches for HCN and HNC in R CrA IRS 7B}
\label{subsubsubsec-ML-RCrAIRS7B}

Following the procedure described in Subsec.~\ref{subsec-finetuning}, we have tackled the analysis of the two spectral lines $J$=3--2 of the main isotopologues HCN and HNC, detected by the APEX telescope in the star forming region R CrA IRS 7B, and the estimates of the excitation temperature and abundance ratios obtained from ML algorithms.

In Tables~\ref{tab_temp-ML-RCrA} and \ref{tab_abundance-ML-RCrA} we present the ML estimates of the excitation temperature and the abundance ratios, respectively, as well as the training and testing coefficients of determination ($\bar R^2_{train}$ and $\bar R^2_{test}$  and the relevant hyperparameters used. In these two tables, we compare the results of the three ML models RF, XGB and MLP when a single line and the two detected lines are considered.

For the temperature prediction in Table~\ref{tab_temp-ML-RCrA} we consider, for the estimates with a single line in this source, the transition $J$=3--2 of HCN  for which the three ML algorithms provide relatively high coefficients of determination. In contrast, the predictions with a single line of HNC have, in general, poor coefficients of determination. In this case, the results attained using the two detected lines are comparable to those accomplished with a single line  HCN 3--2 except for MLP, which coefficients of determination improve  from $\bar R^2_{train}$=0.87 to 0.96 and ${\bar R}^2_{train} - {\bar R}^2_{test}$=0.01. The excitation temperature obtained from the study with two lines are $10\pm0$~K,  $7\pm2$~K and $9\pm1$~K for RF, XGB and MLP, respectively. The XGB prediction is lower than the results of the other two models. Nevertheless,  in accordance with the results obtained in Subsec.~\ref{subsec-finetuning}, the algorithms RF and XGB can be considered in this case the most suitable as well as their higher values of the coefficients of determination.
In fact, the excitation temperatures from RF and XGB are $10\pm0$~K and $7\pm2$~K in agreement with the LTE results obtained for the isotopologues $^{13}$C and $^{15}$N of HCN and HNC although far from those LTE values given for the main isotopologues. 
We expect that this could be corrected when we apply the ML algorithms to a Non-LTE distribution data.

The results of the  HNC/HCN ratios are reported in Table~\ref{tab_abundance-ML-RCrA}. In Subsection \ref{subsec-finetuning}, the HNC/HCN ratios have been estimated using only HNC spectra from Orion KL hot core. For R CrA IRS 7B, initial calculations were performed considering only the HNC $J$=3--2 transition. However, for all three ML models, 
the coefficients of determination do not overcome the value of 0.9. In a subsequent analysis, using the observed spectral pair of HCN and HNC $J$=3--2, we performed new calculations. The results showed that the 
three algorithms provide a ${\bar R}^2_{train}$ close to 1.0 and 
a small difference ${\bar R}^2_{train} - {\bar R}^2_{test}$. Nevertheless, the differences between the models is evident. RF and XGB estimates are rather close,  $0.28\pm0.03$ and $0.29\pm0.08$. These results agree with the LTE ones ranging from 0.30 to 0.39. However, MLP predictions seem unphysical taking a value close to 0.
This discrepancy of MLP with respect to the other algorithms, when using two lines, is due to the fact that MLP can extrapolate and, therefore, the predicted 4-dimensional point lays outside the convex-hull generated by the training sample.

\section{Discussion}
\label{sec-discussion}

In this work, we focus on analysing a complex system rich in astrophysical, astrochemical, and spectroscopic information: the isomeric pair, HCN and HNC. Together, these molecules  encode critical information about the temperatures and evolutionary stages of interstellar sources. This initial focus was driven by considerations about data preprocessing protocols, technical implementation, and the integration of metadata, ensuring a practical framework for training and calibrating the ML models.  
This approach has enabled us to make initial predictions of variables using three ML methods, utilizing a consistent radiative framework for both isomers.

Our approach builds on the diagnostic power of HCN and HNC by constructing a multidimensional dataset through spectral parametrization. This parametrization reveals non-trivial correlations between radiative transfer properties -such as excitation temperatures and HNC/HCN abundance ratios— and observational parameters, including linewidths and intensities, across the $J$=1--0 to $J$=5--4 rotational transitions. In this context, studies have concluded that combined effects, such as  sensitivity, excitation conditions, and the chemistry, can affect the spectral line profiles of different tracers  (e.g., \citealt{Pety2017}).
Recent studies using information-theoretic frameworks~\citep{einig2024} have quantified the value of spectral tracers and their combinations, establishing statistical criteria for constraining physical conditions. Given the complexity of relating spectra to ISM properties, approaches like those and ours are essential for designing targeted observational campaigns and extracting insights from archival molecular line data.

In this present approach, we have tailored three ML models trained on synthetic data. The models were tested using both individual transitions and  combinations of HCN and HNC spectral lines to predict physical parameters, i.e., excitation temperatures and isomeric abundance ratios. This strategy has been implemented with the aim of minimizing the absence of a well-defined function between the input and target spaces and, hence, maximizing the predictive power of the models. Concerning the amount of lines, for single-line predictions, the models produced reasonably accurate results, demonstrating their effectiveness even with limited input data. However, while increasing the number of spectral lines typically can enhance precision, prioritizing transitions or combinations of them -particularly those sensitive to distinct physical parameters— proves more effective and accurate than relying solely on the amount of lines. For instance, the overlap in our parameter datasets distributions (e.g., Fig.~\ref{data-distribHCN-HNC_J5-4}) reveals degeneracies where a single data point can yield multiple physical solutions. This ambiguity is reduced when using two well-chosen transitions, as evidenced by an increase in the $R^2$ coefficient toward unity. Thus, this approach underscores the need to balance computational efficiency with the \ 'informational quality' \ of the spectral lines, which is understood here by their ability to break degeneracies and constrain physical parameters unambiguously. 
In this work, each ML model was trained using 100 iterations of Monte Carlo cross-validation, yielding predictions within minutes. Unlike sampling-based methods, it should be noted that our method does not inherently quantify the uncertainty of the results. Instead, prediction uncertainties were determined as the standard deviation over these iterations. For further studies, an alternative approach could involve sampling Gaussian parameter values at each iteration within ranges defined by their error bars rather than using fixed values. This would provide a more robust uncertainty estimation. 

With regard to the adopted radiative LTE scenario, its selection was a strategic decision aimed at facilitating the data preparation and model. However,  the LTE scenario does not fully explore a number of quantities, such as gas kinetic temperatures and densities, which might be better addressed by adopting non-LTE methods.
These approaches require the application of metadata concerning collisional rates as well as to compare the results obtained from old versus new collisional data files. In this particular case, for a given set of transitions and energy levels of these isomers, the available datasets include the collisional rates of, e.g., HCN and HNC with He spanning a kinetic temperature range of 5–500~K~\citep{Dumouchel2010};  HNC with para-/ortho-H$_2$ for 5–100~K~\citep{Dumouchel2011};  HCN with para-/ortho-H$_2$ for 5–100~K~\citep{Vera2014}; and HCN/HNC with para-/ortho-H$_2$  covering a kinetic temperature range of 10–500~K~\citep{HernandezVera2017}. Thus, the line analysis and non-LTE models require a more detailed examination during the data preparation stage. Notwithstanding these constraints, our results demonstrate
the LTE framework enables the training and validation of ML models for predicting excitation temperatures and HNC/HCN abundance ratios. This represents an essential step towards refining the method into a more comprehensive and unbiased approach, capable of incorporating other molecules, astrophysical sources, and varying physical conditions. A comprehensive study incorporating new observed data, non-LTE analysis, and ML techniques is planned in a forthcoming work, where
the outcomes obtained with different datasets of collisional rates will be assessed.

\section{Concluding remarks}
\label{sec4}

The main contributions of this work are the followings: i)  For the first time, a data cloud has been explored and generated based on Gaussian fit parameters derived from the spectral lines of HCN and HNC under LTE conditions. This data cloud serves as a database that links physical properties, such as excitation temperatures and HNC/HCN ratios, to line parameters, thereby  enhancing significantly the capability of ML models to perform unsupervised predictions of these physical properties;  
ii) we propose a bottom-up approach that, contrary to the top-down mainstream approaches, starts with the simulation of a set of spectral data at different physical conditions to interpret line observations of interstellar regions and to estimate of their physical conditions; and iii) this new approach presents the potential relevance to unravel hidden ISM conditions with the use of ML approaches. This approach has been proven rather precise, according to the low standard deviations of the ML predictions, and rather fast, i.e., taking barely a  few minutes on a modern laptop for the complete optimization procedure (training, testing and solving the target quantities) once the spectral simulations are obtained. 
Although this approach has been validated with observations of the isomers HCN and HNC carried out with a single dish radio antenna, 
 our method should also work for interferometers and space-based observatories.  In fact, this approach has a broader scope since it aims at the observations of higher rotational and vibrational transitions as well as any other molecule.

On the one hand, a comprehensive set of spectral data simulations have been reproduced for different values of the physical conditions, collecting the information of their profiles in the input variables (line intensity and line width) used in ML approaches. 
In the course of this study we have become aware that
it can be beneficial to carry out a preliminary analysis of the spectral lines under the physical conditions.
The preliminary analysis of the data cloud distribution has been used to rule out those spectral lines with a strong superposition in the data cloud distribution. This discrimination makes more efficient and precise  the ML prediction for determining the target variables (the excitation temperature and the abundance ratio HNC/HCN).
This step can also help us to choose the interval of physical quantities to be explored improving the efficiency and saving the computational cost as well as delimiting the set of data considered in the ML estimates of the ISM physical conditions. 

On the other hand, three  Machine Learning models MLP, RF and XGB have been trained, tested and compared. First  an assessment of the ML algorithms has been carried out using the semiempirical data of Orion KL hot core. For this test, the excitation temperature was estimated with the identified transitions $J$=5--4 of HCN and $J$=4–-3 of HNC and the  ratios with $J$=4–-3 and $J$=3–-2 of HNC. The three models gave rise to comparable results and with a good precision  except for the MLP model because, when two lines are considered, this tends to extrapolate to values out of the set of training data. Hence, for this particular case, the most suitable algorithms are RF and XGB  providing similar predictions and close values for $\bar R^2_{train}$ and $\bar R^2_{test}$.  The RF and XGB models resulted in excitation temperatures and ratios of $90\pm1$~K and $89\pm1$~K, and $0.81\pm0.02$ and $0.82\pm0.02$, respectively.
 The LTE model provided a $T_{\rm exc} \thickapprox$ 90~K and $N$(HNC)/$N$(HCN)$\thickapprox$~0.8. This result drove us to address the study of a more realistic source, the star forming region R CrA IRS 7B, in which two spectral lines $J$=3–2 of the isotopologues of HCN and HNC were detected by the APEX telescope. For this source and the detected lines, we found that, in general, the estimates of the excitation temperature and ratios obtained from the  three ML algorithms are in agreement with the LTE results  with the exception of 
 the MLP ratio prediction, which goes down drastically to the unphysical value of zero because of its tendency to extrapolate to values out of the set of training data when the two detected lines are considered. For the analysis of the two detected lines of main isotopologues,  RF and XGB models are considered the most suitable according to the results obtained  in Subsec.~\ref{subsec-finetuning}. The predictions of RF and XGB resulted in  excitation temperatures of $10\pm0$~K and $7\pm2$~K, and ratios of $0.28\pm0.03$ and $0.29\pm0.08$, respectively. LTE approach determined an excitation temperature from $11\pm3$ to $14\pm2$~K for the $^{13}$C and $^{15}$N isotopologues, in agreement with RF  and XGB models, and $26\pm5$~K and $27\pm5$~K for the main isotopologue, although it is argued in the previous section~\ref{subsec-RCrA} that Non-LTE calculations should be considered in this case. The LTE result for the HNC/HCN ratios is between $0.30\pm0.03$ and $0.39\pm0.25$, also in accordance to  RF and XGB algorithms.

In addition, this work could be significant for determining the physical conditions of surveys where a single spectral line is identified.
For the semiempirical data of Orion KL hot core, the only use of the transition $J$=5--4 of HCN for estimating the excitation temperature and of the transition $J$=4--3 of HNC for the estimate of the ratio resulted in similar results considering the two transitions. 
For the source R CrA IRS 7B, the estimates of the excitation temperature with the detected line $J$=3--2 of HCN are also similar to those obtained with the two lines. However, this fact does not happen when the  abundance ratio is predicted with only the line $J$=3–-2 of HNC. Therefore, the application to only one line reinforces the idea that the determination of the physical parameters using only one detected line is possible although it would be advisable to carry out a further analysis, i.e., considering different algorithms and spectral lines.  This research opens a new expectation as to determine the physical conditions from the location of the point in the data cloud for a given identified spectral line.

In the near future, we are going to carry out the following studies:
\begin{itemize}
\item We will apply this model to determine gas kinetic temperatures using non-LTE assumptions.
\item We will explore the effect of the generalized Gaussian distribution for a better shape of the simulated spectral lines considering an extra parameter apart from $I$  and $\Delta \nu$. It is expected that this extra parameter will help us to take advantage of the information of the spectral line profiles allowing us to determine more physical parameters and with a better accuracy. 
\item The associated natural line is primarily determined by Doppler shifts that reflect the radial velocities of the emitting atoms and molecules. In future works, we will also incorporate Lorentzian, Voigt, and asymmetric line profiles to examine bulk gas affected by the presence of jets and molecular outflows. 
\item We will examine the results obtained from a grid considering, as input variables, different values of the column density of HCN, the source's size and other factors that could affect the spectral line profiles.  
\item  We will compare the ML results obtained using new detected spectral lines of HCN and HNC, involving higher excited rovibrational states, and of other molecular species.
\item We will contemplate the application of this approach to surveys of the Earth atmosphere.
\end{itemize}

\begin{acknowledgements}

 We would like to thank the anonymous reviewer for their valuable and careful revision of the manuscript. We are in debt for the strategies suggested in the report.
We wish to acknowledge useful discussions with Natalia Inostroza, Diego Mardones, Manuel Merello and Leonardo Bronfman. E.M. acknowledges support under the grant "María Zambrano" from the University of Huelva funded by the Spanish Ministry of Universities and the "European Union NextGenerationEU". 
P.D. acknowledges financial support from “Junta de
Andalucía” through ”Programa Operativo FEDER de Andalucía 2014-2020 (PAIDI)” under the project P20 00764, and E.M. and M.C. from “Junta de
Andalucía” through ”Programa Operativo FEDER de Andalucía 2021-2027 (PAIDI)” under the project EPIT1462023.
This project has also received funding from the European Union's Horizon 2020 research and innovation program under Marie Sklodowska-Curie grant agreement No. 872081, grants PID2020-119478GB-I00 (A.P.) and PID2022-136228NB-C21 (M.C.) funded by MCIN/AEI/10.13039/501100011033, and, as appropriate, by "ERDF A way of making Europe", the "European Union", or the "European Union NextGenerationEU/PRTR". This work is also supported by the Consejería de Transformación Económica, Industria, Conocimiento y Universidades, Junta de Andalucía and European Regional Development Fund (ERDF 2014-2020) PY2000764.
This work is based on data acquired with the Atacama Pathfinder Experiment (APEX) under programs O-090.F-9317A-2012,  O-094.F-9321A-2014,   E-0104.C-0033A-2019,  O-0107.F-9303A-2021. APEX is a collaboration between the Max-Planck-Institut für Radioastronomie, the European Southern Observatory, and the Onsala Space Observatory. 

\end{acknowledgements}

   \bibliographystyle{aa} 
   \bibliography{aa-MLversion} 

\begin{appendix}

\section{Data distribution of the profiles of the simulated spectral lines of HCN and HNC.}
\label{app-charts}

\begin{figure*}
  \begin{center}
    \setlength{\unitlength}{1pt} 
    \includegraphics[width = 9.1cm]{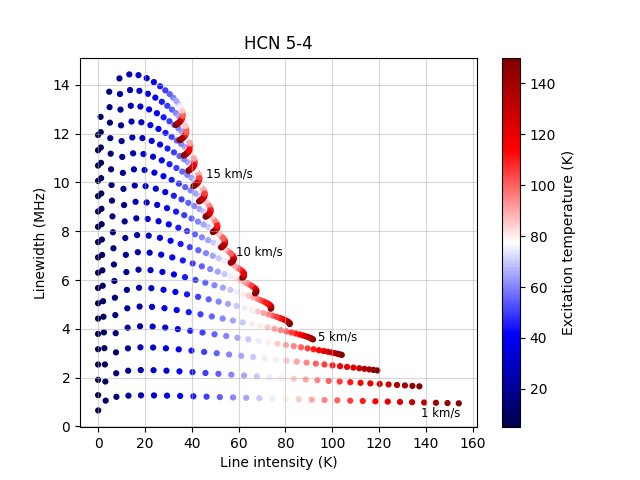}\includegraphics[width = 9.1cm]{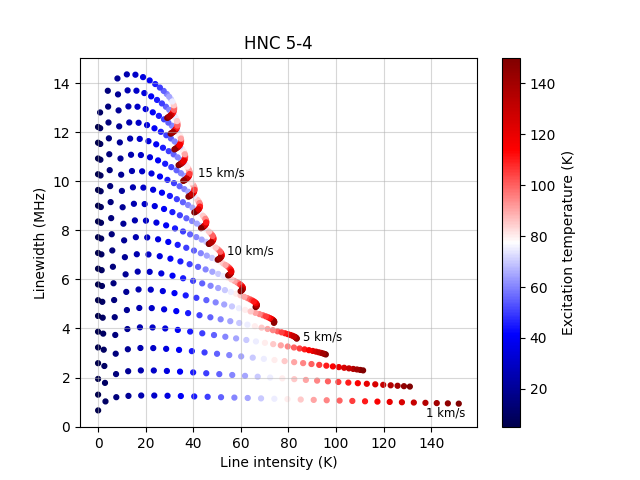}
\includegraphics[width = 9.1cm]{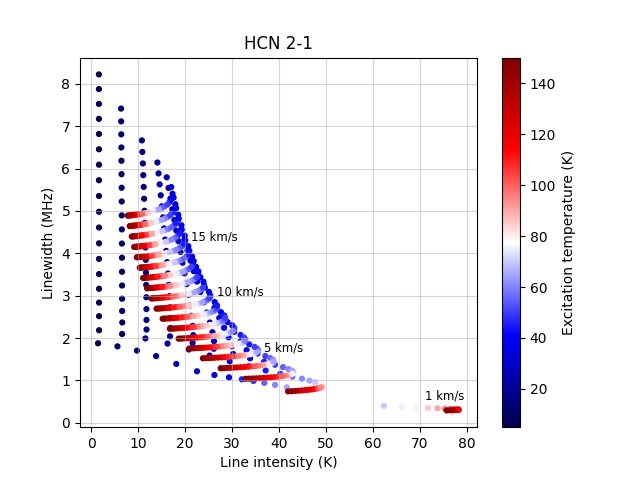}\includegraphics[width = 9.1cm]{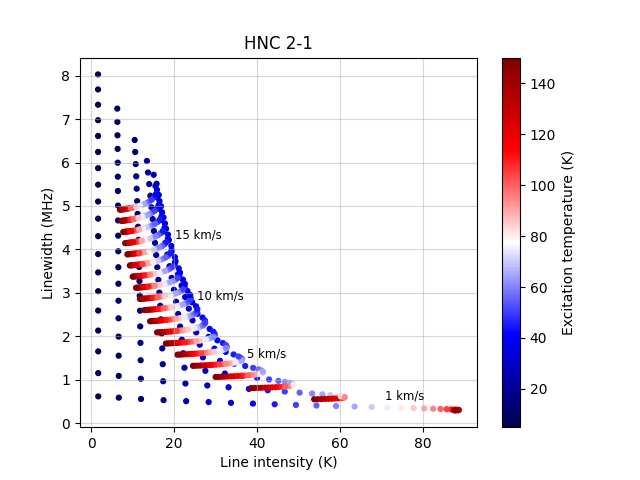}
    \caption{Data distributions of the profiles of the spectral lines $J$=5-4 and $J$=2-1 of HCN and HNC parameterized with the  variables ($I$,$\Delta \nu$). The upper and lower rows show the data distributions of J=5-4 and J=2-1, respectively, for HCN (left panels) and HNC (right panels).
    The relative abundance HNC/HCN and the source's size are fixed to 0.8 and 5 arcsec, respectively. 
    As the data are distributed by branches of points, corresponding to the steps  of $\Delta V_{\rm sys}$, only those with $\Delta V_{\rm sys}$=1, 5, 10 and 15~km/s are showcased for the sake of clarity.}
    \label{data-distribHCN-HNC_J5-4}
  \end{center}
\end{figure*}

\begin{figure*}
  \begin{center}
    \setlength{\unitlength}{1pt} 
    \includegraphics[width = 8.1cm]{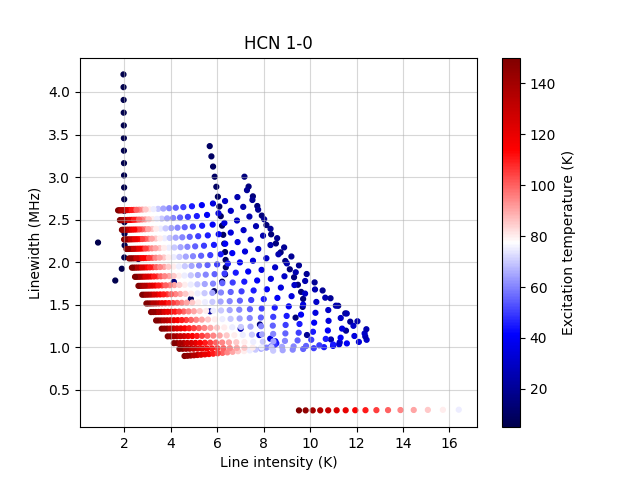}\includegraphics[width = 8.1cm]{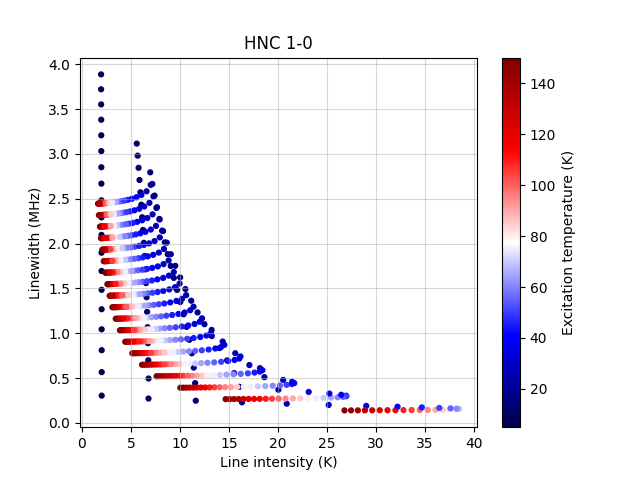}
    \includegraphics[width = 8.1cm]{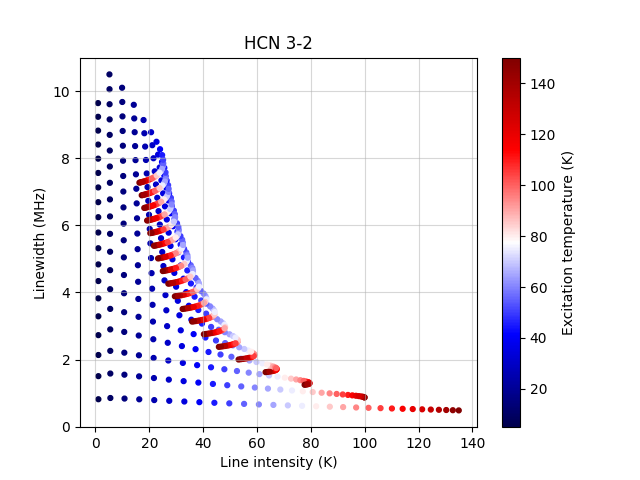}\includegraphics[width = 8.1cm]{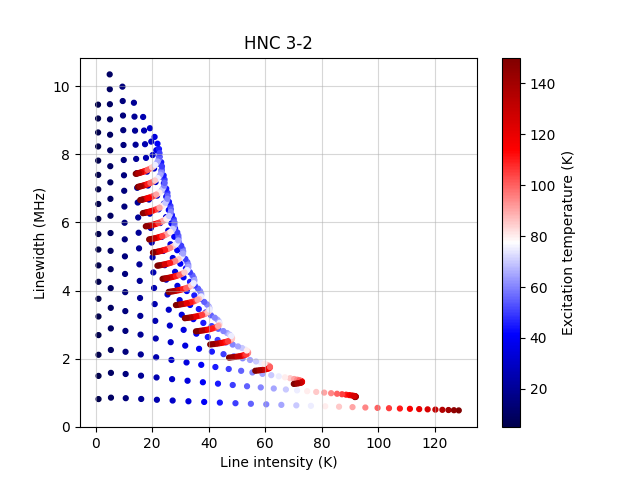}
    \includegraphics[width = 8.1cm]{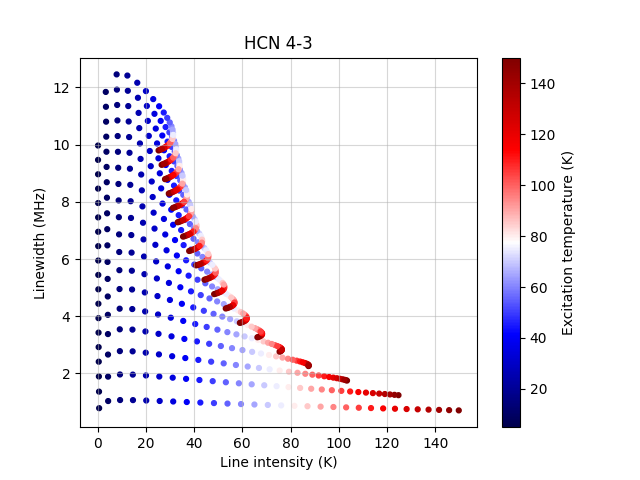}\includegraphics[width = 8.1cm]{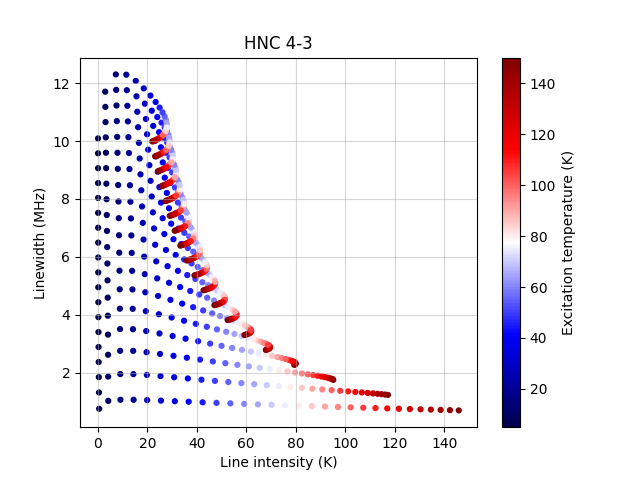}
    \caption{Data distributions of the profiles of the spectral lines J=1--0, $J$=3--2 and $J$=4--3 of HCN and HNC parameterized with the  variables ($I$,$\Delta \nu$). The upper, middle and lower rows show the data distributions for $J$=1--0, $J$=3--2 and $J$=4--3, respectively. Left panels belong to HCN and right panels to HNC transitions.
    The relative abundance HNC/HCN and the source's size are fixed to 0.8 and 5 arcsec, respectively.}
    \label{data-distribHCN-HNC_app}
  \end{center}
\end{figure*}

In Fig.~\ref{data-distribHCN-HNC_app}, other temperature charts of the data are shown for the profiles of the spectral transitions $J$=1--0, $J$=3--2  and $J$=4--3 of HCN and HNC. As in Fig.~\ref{data-distribHCN-HNC_J5-4}, the  HNC/HCN ratio and the source's size are fixed to 0.8 and 5 arcsec, respectively. These displays are exhibited to support the statement that the distribution of data are more scattered when the higher excited states are involved in the transition lines, up least up to $J$=5.

\begin{table*}  
\caption{Performance metrics and temperature predictions for Orion KL hot core from the RF, XGB, and MLP models. \label{tab_temp-ML}}
\centering
\begin{tabular}{ccccccc}
\multicolumn{7}{c}{RF} \\ \hline
 Spectral lines& $\bar R^2_{train}$ & $\bar R^2_{test}$ & $\bar T_{pred}\,\, (K) $& $\mathrm{Std}_{pred}\,\,(K)$ & depth\\
 \hline
 HCN 5--4 & 0.99     &  0.99  & 90 & 0 & 20\\
 HNC 4--3 & 0.62  & 0.61 & 100  & 2 & 5\\
 \mbox{HCN 5--4}  \mbox{HNC 4--3} & 0.999  & 0.998   & 90 & 1 & 20 \\
 \hline
 \multicolumn{7}{c}{} \\
 \multicolumn{7}{c}{XGB} \\ \hline
 Spectral lines& $\bar R^2_{train}$ & $\bar R^2_{test}$ & $\bar T_{pred}\,\, (K) $& $\mathrm{Std}_{pred}\,\,(K)$ & depth & learning rate\\
 \hline
 HCN 5--4 &  0.99    & 0.99   & 90 & 1  & 10 & 0.01\\
 HNC 4--3 & 0.60   & 0.58  & 91   & 1 & 5 & 0.003 \\
 \mbox{HCN 5--4}  \mbox{HNC 4--3} & 0.999  & 0.998   & 89 & 1 & 10 & 0.1\\
 \hline
 \multicolumn{7}{c}{} \\
 \multicolumn{7}{c}{MLP} \\ \hline
Spectral lines & $\bar R^2_{train}$ & $\bar R^2_{test}$ & $\bar T_{pred}\,\, (K) $& $\mathrm{Std}_{pred}\,\,(K)$\\
 \hline
 HCN 5--4 & 0.96  & 0.96  & 93 & 3 \\
 HNC 4--3 & 0.62 & 0.60 & 105  & 6 \\
  \mbox{HCN 5--4}  \mbox{HNC 4--3} & 0.99 & 0.99 & 68 & 8 \\
 \hline
\end{tabular}
\tablefoot{The mean coefficients of determination are given by $\bar R^2_{train}$ and $\bar R^2_{test}$ for the RF, XGB, MLP models trained with different combinations of spectral lines as well as the mean values and standard deviations of the temperature prediction for Orion KL hot core source.  The adjusted relevant hyperparameters of RF and XGB are also reported. The hyperparameters of MLP are described in Sect.\ref{subsec-ML}.}
\end{table*}

\begin{table*}  
\caption{Performance metrics and ratios predictions of HNC/HCN for the Orion KL hot core from the RF, XGB, and MLP models.}\label{tab_abundance-ML}
\centering
\begin{tabular}{ccccccc}
\multicolumn{7}{c}{RF} \\ \hline  
 Spectral lines& $\bar R^2_{train}$ & $\bar R^2_{test}$ & $\!\!\overline{\mbox{HNC/HCN}}$ & $\mathrm{Std}_{pred}$ & depth\\
 \hline
 HNC 3--2 & 0.77  & 0.74  & 0.83  & 0.02 &  6\\
 HNC 4--3 & 0.66   & 0.63  & 0.74    &0.02 & 5\\
 \mbox{HNC 3--2}  \mbox{HNC 4--3} & 0.86 & 0.84  & 0.81  & 0.02 & 6\\
 \hline
\multicolumn{7}{c}{} \\
\multicolumn{7}{c}{XGB} \\ \hline
 Spectral lines& $\bar R^2_{train}$ & $\bar R^2_{test}$ & $\!\! \overline{\mbox{HNC/HCN}}$& $\mathrm{Std}_{pred}$ & depth & learning rate\\
 \hline
 HNC 3--2 & 0.81  & 0.79  & 0.82  & 0.02 & 2 & 0.05 \\
 HNC 4--3 & 0.72  & 0.69 & 0.80  & 0.02 & 2 & 0.05\\
 \mbox{HNC 3--2}  \mbox{HNC 4--3} & 0.96 &0.94  &0.82  &0.02 & 3 & 0.1 \\
 \hline
 \multicolumn{7}{c}{} \\
 \multicolumn{7}{c}{MLP} \\ \hline
 Spectral lines& $\bar R^2_{train}$ & $\bar R^2_{test}$ & $\!\! \overline{\mbox{HNC/HCN}} $& $\mathrm{Std}_{pred}$\\
 \hline
 HNC 3--2 & 0.77  & 0.75  & 0.82  &0.04 \\
 HNC 4--3 & 0.73  & 0.71 & 0.80  & 0.04  \\
 \mbox{HNC 3--2}  \mbox{HNC 4--3} & 0.96 & 0.96  &0.72  &0.04 \\
 \hline
\end{tabular}
\tablefoot{Likewise Table~\ref{tab_temp-ML},  mean values of the coefficients of determination ($\bar R^2_{train}$ and $\bar R^2_{test}$) are provided for the three models.}
\end{table*}

\begin{table*}  
\caption{Performance metrics and temperature predictions for the  cold
source R CrA IRS 7B from the RF, XGB, and MLP models.}\label{tab_temp-ML-RCrA}
\centering
\begin{tabular}{ccccccc}
\multicolumn{7}{c}{RF} \\ \hline
 Spectral lines& $\bar R^2_{train}$ & $\bar R^2_{test}$ & $\bar T_{pred}\,\, (K) $& $\mathrm{Std}_{pred}\,\,(K)$ & depth\\
 \hline
 HCN 3--2 & 0.99  & 0.99 & 10  & 0 & 20\\
 \mbox{HCN 3--2}  \mbox{HNC 3--2} & 0.99  & 0.98   & 10 & 0 & 20 \\
 \hline
\multicolumn{7}{c}{}\\ 
 \multicolumn{7}{c}{XGB} \\ \hline
Spectral lines & $\bar R^2_{train}$ & $\bar R^2_{test}$ & $\bar T_{pred}\,\, (K) $& $\mathrm{Std}_{pred}\,\,(K)$ & depth & learning rate\\
 \hline
 HCN 3--2 & 0.99  & 0.99 & 5  & 1 & 6 & 0.05 \\
 \mbox{HCN 3--2}  \mbox{HNC 3--2} & 0.99  & 0.98   & 7 & 2 & 5 & 0.05 \\
 \hline
\multicolumn{7}{c}{}\\ 
\multicolumn{7}{c}{MLP} \\ \hline
Spectral lines & $\bar R^2_{train}$ & $\bar R^2_{test}$ & $\bar T_{pred}\,\, (K) $& $\mathrm{Std}_{pred}\,\,(K)$\\
 \hline
HCN 3--2 & 0.87  & 0.86 & 9  & 1 \\
 \mbox{HCN 3--2}  \mbox{HNC 3--2} & 0.96  & 0.95   & 9 & 1 \\
 \hline
\end{tabular}
\tablefoot{Likewise Table~\ref{tab_temp-ML},  mean values of the coefficients of determination ($\bar R^2_{train}$ and $\bar R^2_{test}$) are provided for the three models.}
\end{table*}

\begin{table*}  
\caption{Performance metrics and ratios predictions of HNC/HCN for the cold
source R CrA IRS 7B  from the RF, XGB, and MLP models.}\label{tab_abundance-ML-RCrA}
\centering
\begin{tabular}{ccccccc}
\multicolumn{7}{c}{RF} \\ \hline  
 Spectral lines& $\bar R^2_{train}$ & $\bar R^2_{test}$ & $\!\!\overline{\mbox{HNC/HCN}}$ & $\mathrm{Std}_{pred}$ & depth\\
 \hline
 HNC 3--2 & 0.80  & 0.77 & 0.11  & 0.004 & 6\\
 \mbox{HCN 3--2}  \mbox{HNC 3--2} & 0.99  & 0.98   & 0.28 & 0.03 & 12 \\
  \hline
\multicolumn{7}{c}{}\\ 
\multicolumn{7}{c}{XGB} \\ \hline
Spectral lines & $\bar R^2_{train}$ & $\bar R^2_{test}$ & $\!\! \overline{\mbox{HNC/HCN}}$& $\mathrm{Std}_{pred}$ & depth & learning rate\\
 \hline
HNC 3--2 & 0.86  & 0.84 & 0.11  & 0.01 & 3 & 0.03 \\
 \mbox{HCN 3--2}  \mbox{HNC 3--2} & 0.99  & 0.98   & 0.29 & 0.08 & 5 & 0.05 \\
 \hline
 \multicolumn{7}{c}{}\\ 
 \multicolumn{7}{c}{MLP} \\ \hline
Spectral lines & $\bar R^2_{train}$ & $\bar R^2_{test}$ & $\!\! \overline{\mbox{HNC/HCN}} $& $\mathrm{Std}_{pred}$\\
 \hline
HNC 3--2 & 0.77  & 0.75 & 0.34  & 0.1 \\
 \mbox{HCN 3--2}  \mbox{HNC 3--2} & 0.99  & 0.99   & 0.004 & 0.008 \\
 \hline
\end{tabular}
\tablefoot{Likewise Table~\ref{tab_temp-ML},  mean values of the coefficients of determination ($\bar R^2_{train}$ and $\bar R^2_{test}$) are provided for the three models.}
\end{table*}

\begin{figure}
\centering
\includegraphics[scale=0.35]{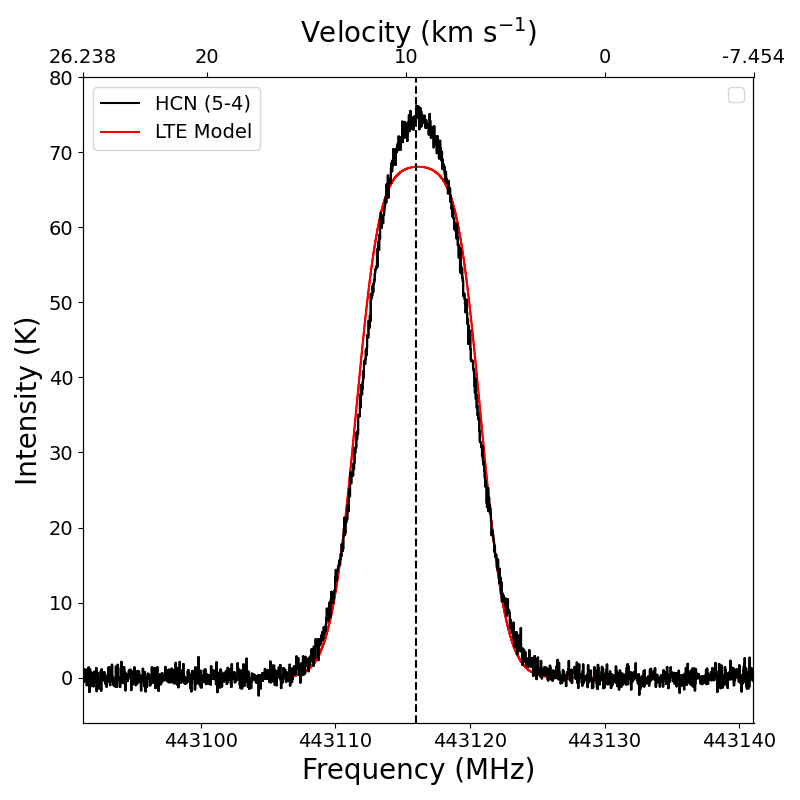}\\
\includegraphics[scale=0.35]{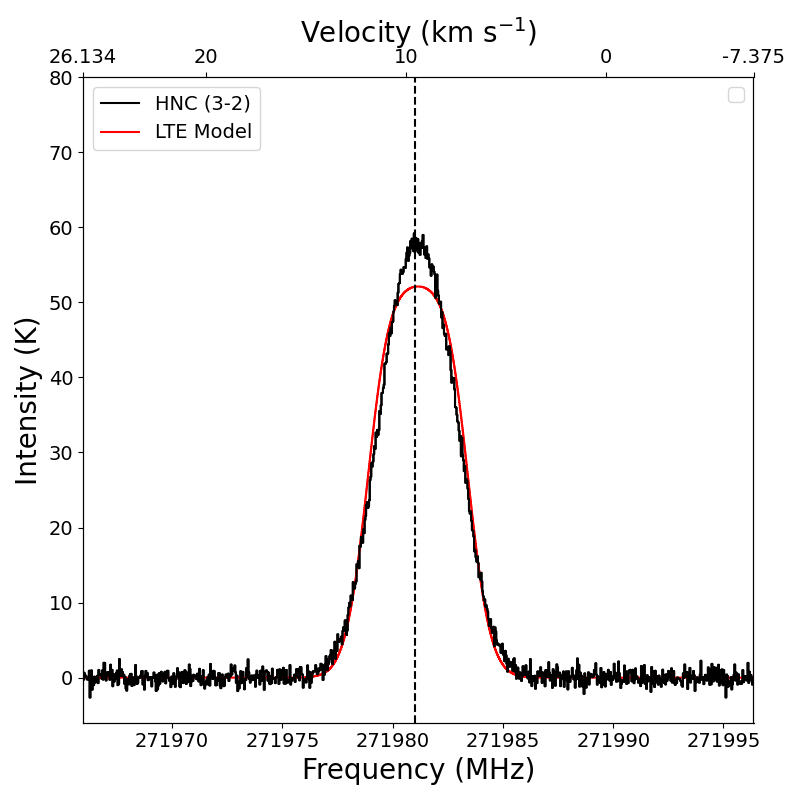}\\
\includegraphics[scale=0.35]{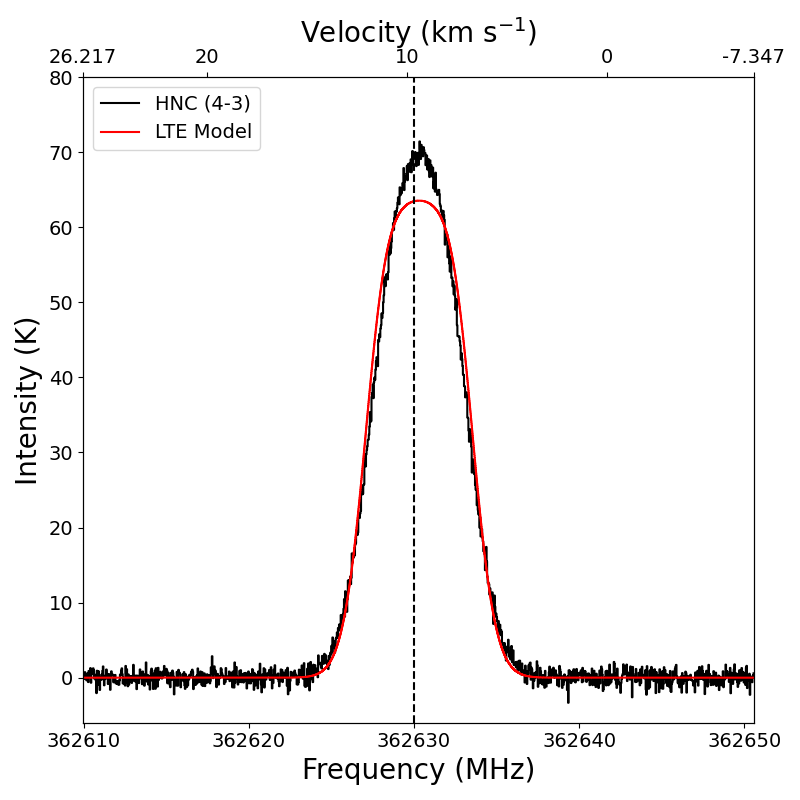}\\
\caption{Spectra of the HCN (5--4), HNC (3--2), and HNC (4--3) transitions toward Orion KL ($V_{lsr} \thickapprox$  9.4 km s$^{-1}$) combined with LTE-MCMC models representing the conditions of a hot core with an excitation temperature of 90 K and $N$(HNC)/$N$(HCN)=0.8, calculated with a fixed value of $N$(HCN) $\simeq 1 \times 10^{15}$ cm$^{-2}$.}
\label{fig:hcn-hnc-ori}
\end{figure}

\begin{figure}
  \begin{center}
    \setlength{\unitlength}{1pt} 
    \includegraphics[width = 9.1cm]{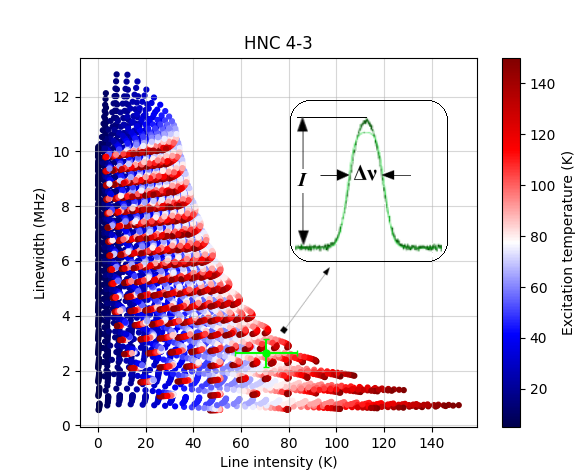}\\
    \includegraphics[width = 9.1cm]{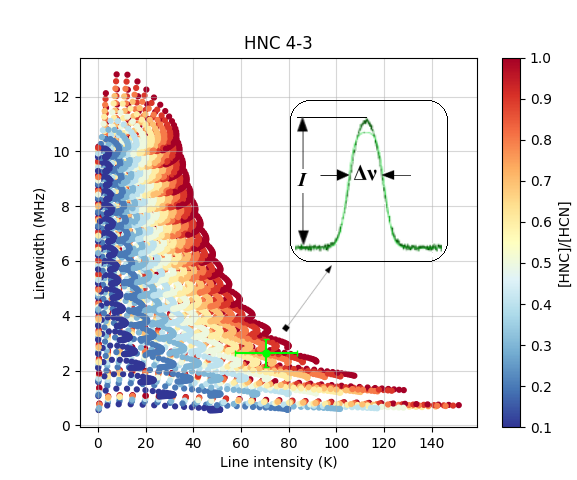}
    \caption{Data distribution of the profiles of the simulated spectral line $J$=4-3 of HNC for the grid of physical parameters values. The top display shows the temperature mapping of the data while the bottom one shows the relative abundance mapping. Both panels depict the HNC $J$=4-3 line transition, from which the line intensity and linewidth, along with their associated error bars, are obtained for the semiempirical data of Orion KL. These values are indicated by a green dot in each panel.} 
    \label{data-distribHNC_J4-3}
  \end{center}
\end{figure}

\begin{figure*}
\centering
\includegraphics[scale=0.35]{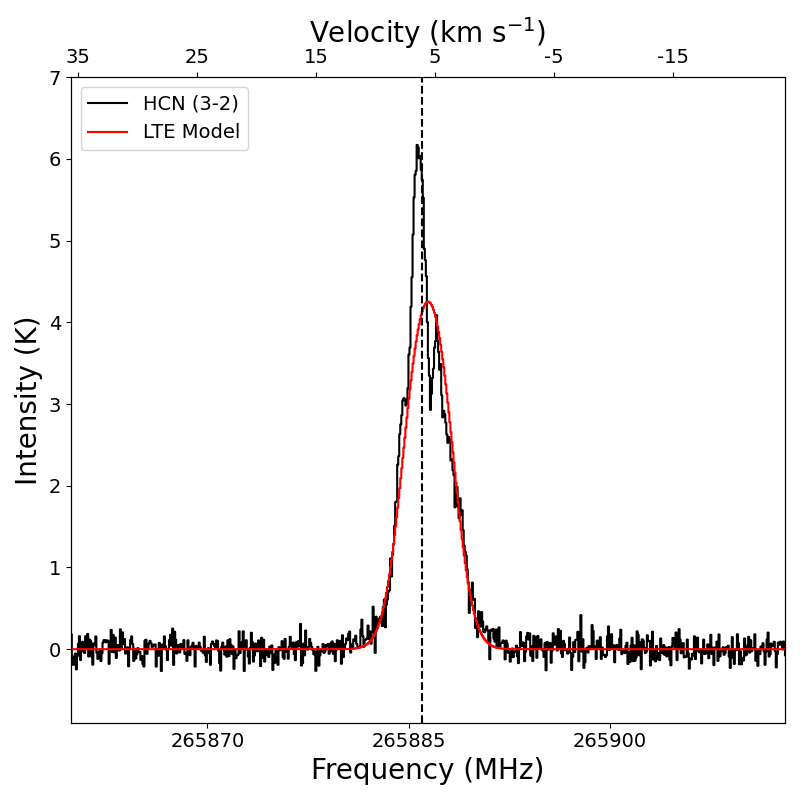}     \includegraphics[scale=0.35]{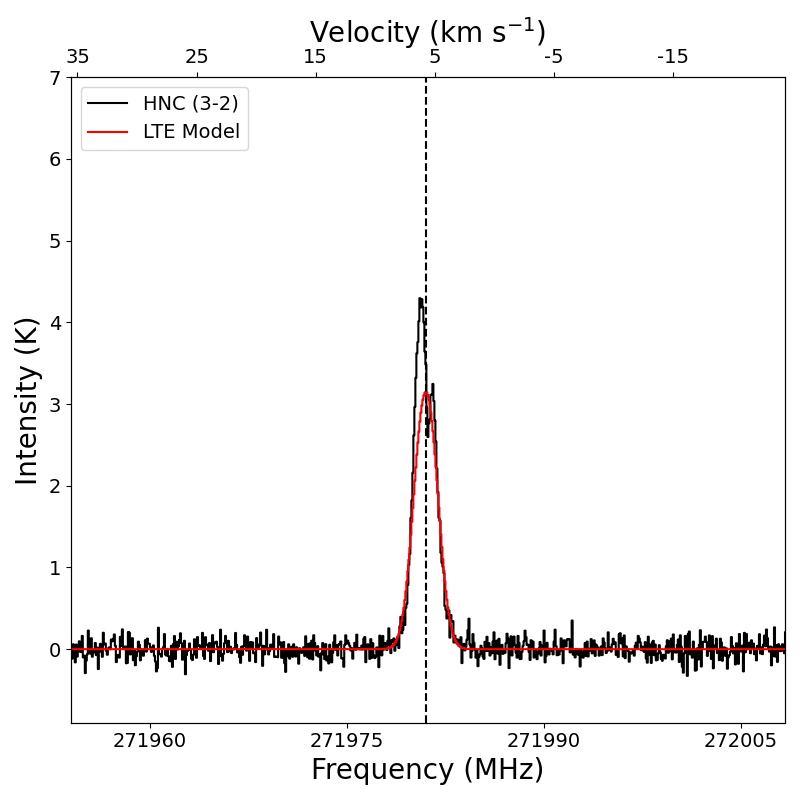}\\
\includegraphics[scale=0.35]{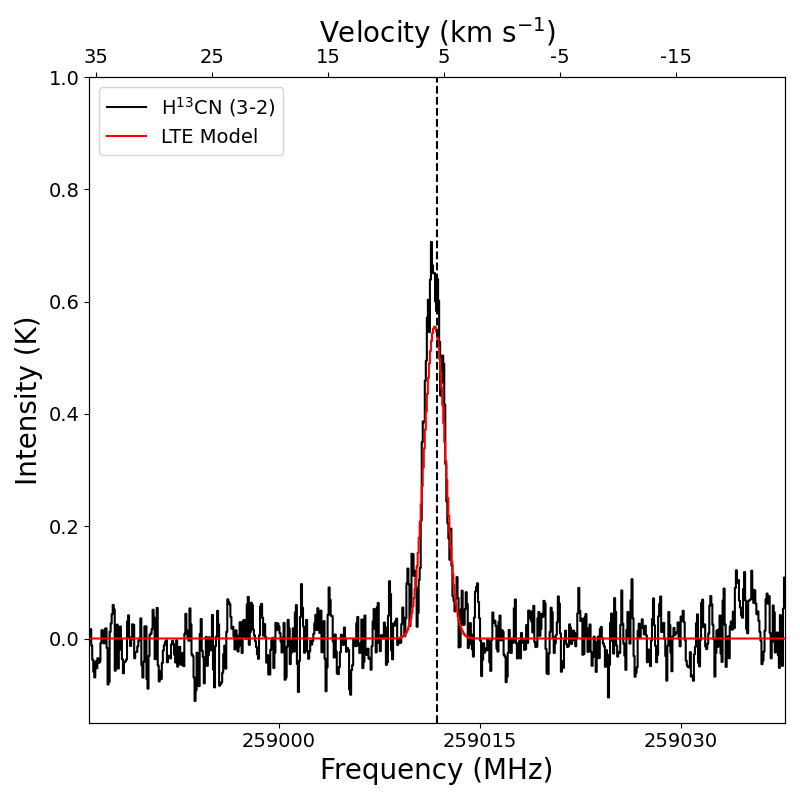}   \includegraphics[scale=0.35]{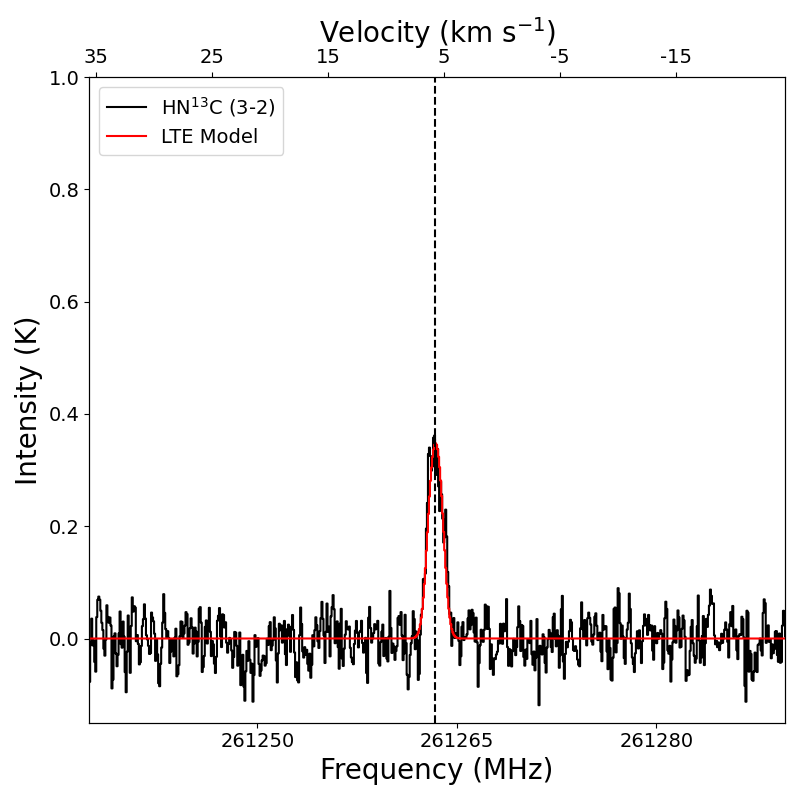}\\
\vspace{0.5cm}
\includegraphics[scale=0.35]{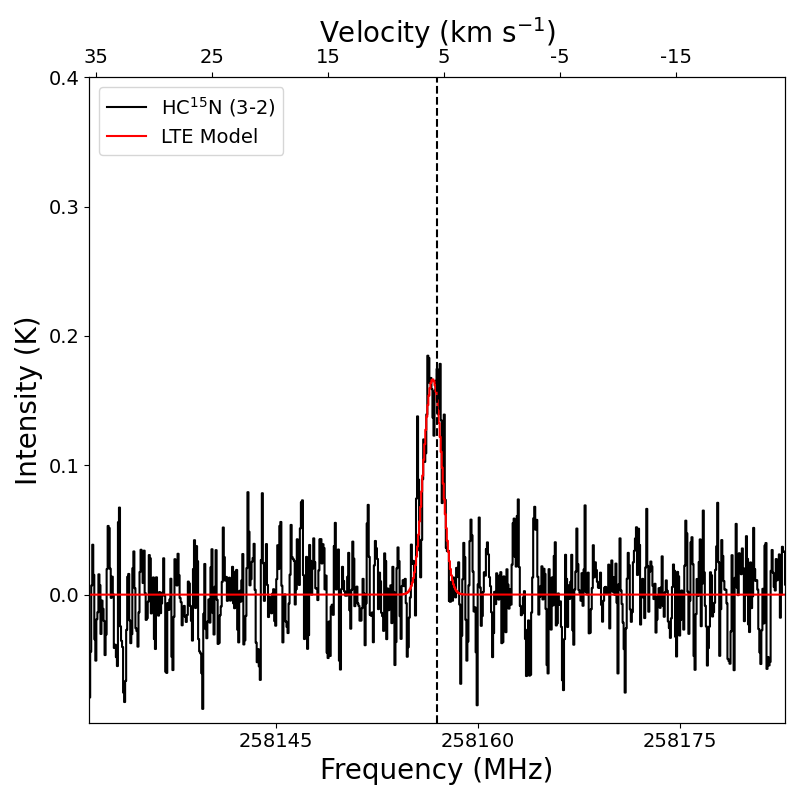}   \includegraphics[scale=0.35]{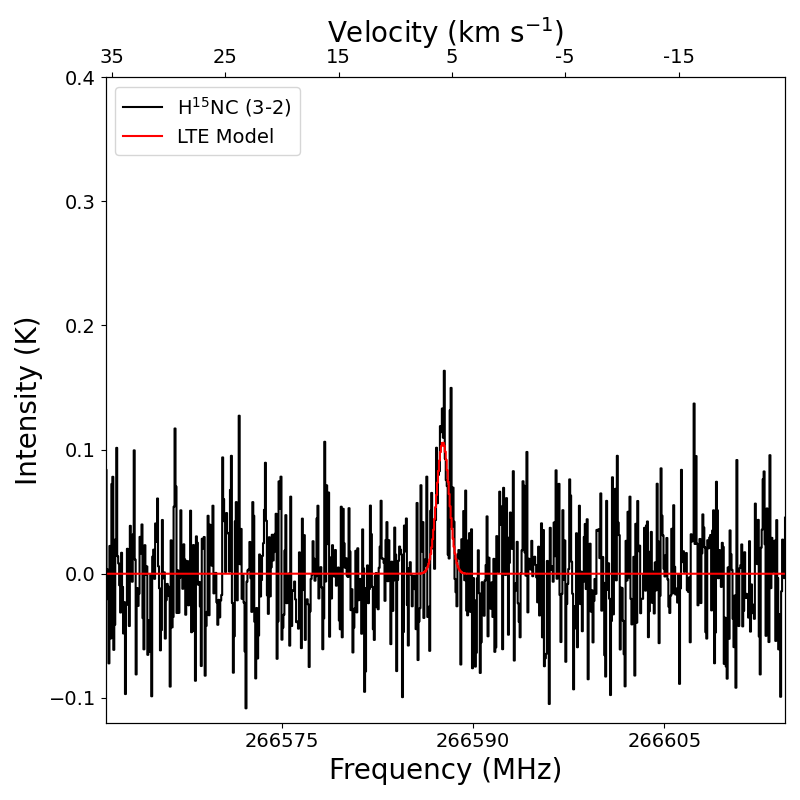}\\
\caption{APEX-detected spectral lines of HCN and HNC isotopologues towards R CrA IRS 7B (black histogram), with corresponding LTE-MCMC models shown as solid red lines. Upper panels: HCN and HNC $J$=3-2 transitions, exhibiting excitation temperatures ($T_{\text{exc}}$) of 26(5)~K and 27(5)~K, respectively, and a column density ratio $N(\text{HNC})/N(\text{HCN}) \thickapprox 0.30$. Middle panels: H$^{13}$CN and HN$^{13}$C $J$=3-2, with $T_{\text{exc}}$ of 11(3)~K and 12(2)~K, respectively, and $N(\text{HN}^{13}\text{C})/N(\text{H}^{13}\text{CN}) \thickapprox 0.38$. Lower panels: HC$^{15}$N and H$^{15}$NC $J$=3-2, with $T_{\text{exc}}$ of 11(4)~K and 14(2)~K, respectively, and $N(\text{HN}^{15}\text{C})/N(\text{HC}^{15}\text{N}) \thickapprox$ 0.39.}
\label{fig:hcn-hnc-rcr}
\end{figure*}

\end{appendix}

\end{document}